\documentclass{aa}
\input{psfig}
\def\tvi(#1,#2){\vrule height #1pt depth #2pt width 0pt}
\def\p{\partial}
\def\e{{\rm e}}
\def\d{{\rm d}}
\def\ie{i.e. }
\def\eg{e.g. }
\def\etal{et al. }
\def\rs{{r}_{\rm sh}}
\def\M{{\cal M}}
\def\growth{\sigma}
\def\grwdth{h}
\def\pttwo{\alpha}
\def\seven{\kappa}

\def\KHi{KHi }
\def\RTi{RTi }
\def\vff{v_{\rm ff}}
\def\vmax{\varpi_{\rm max}}
\def\MA{{\hat{\cal A}_i}}
\def\nix{\discretionary{}{}{}}
\def\rA{r_{\rm A}}
\newlength{\largeur}
\newlength{\saut}
\def\marge#1{
\setlength{\largeur}{\columnwidth}
\addtolength{\largeur}{-#1}
\setlength{\saut}{0.5\largeur}\hspace*{\saut}} \def\picture #1 by #2
(#3){
\marge{#1} \vbox to #2{
\hrule width #1 height 0pt depth 0pt
\vfill
\special{picture #3}}}
\def\scaledpicture #1 by #2 (#3 scaled #4){{
\dimen0=#1 \dimen1=#2
\divide\dimen0 by 1000 \multiply\dimen0 by #4 \divide\dimen1 by 1000
\multiply\dimen1 by #4 \picture \dimen0 by \dimen1 (#3 scaled #4)}}
\begin{document}

\thesaurus{06
(02.01.2;
02.08.1;
02.09.1;
02.19.1;
08.02.1;
13.25.5)}

\title{An analytic study of Bondi--Hoyle--Lyttleton accretion}
\subtitle{II. Local stability analysis}
\author{T. Foglizzo\inst{1}\thanks{e-mail: {\tt foglizzo@cea.fr}}
\and M. Ruffert\inst{2}\thanks{e-mail: {\tt m.ruffert@ed.ac.uk}} }

\offprints{T.~Foglizzo}

\institute {Service d'Astrophysique, CEA/DSM/DAPNIA, CE-Saclay, 91191
Gif-sur-Yvette, France \and 
Dept. of Maths. \& Stats., Univ. of Edinburgh, Edinburgh EH9 3JZ, 
Great Britain}

\date{Received ; accepted }
\titlerunning{II. Local stability analysis}
\maketitle
%\maintitlerunninghead{...}
%\authorrunninghead{}

\begin{abstract}
The adiabatic shock produced by a compact object moving supersonically 
relative to a gas with uniform entropy and no vorticity is a source of 
entropy gradients and vorticity. We investigate these analytically.
The non--axisymmetric Rayleigh--Taylor and axisym\-metric Kel\-vin--Helmholtz 
linear instabilities are potential sources of destabilization of the 
subsonic accretion flow after the shock. A local Lagrangian approach is used 
in order to evaluate the efficiency of these linear instabilities. 
However, the conditions required for such a WKB type 
approximation are fulfilled only marginally: a quantitative estimate of 
their local growth rate integrated along a flow line shows that their 
growth time is at best comparable to the time needed for advection onto the 
accretor, even at high Mach number and for a small accretor size.  Despite 
this apparently low efficiency, several features of these mechanisms 
qualitatively match those observed in numerical simulations: in a gas with 
uniform entropy, the instability occurs only for supersonic accretors. It is 
nonaxisymmetric, and begins close to the accretor in the 
equatorial region perpendicular to the symmetry axis. The mechanism is more 
efficient for a small, highly supersonic accretor, and also if the shock is 
detached.

We also show by a 3--D numerical simulation an example of unstable
accretion of a subsonic flow with non--uniform entropy at infinity.
This instability is qualitatively similar to the one observed in
3--D simulations of the Bondi--Hoyle--Lyttleton flow, although it involves
neither a bow shock nor an accretion line.


\end{abstract}

\section{Introduction}

The instability of the supersonic
axisymmetric Bondi--Hoyle--Lyttleton (hereafter BHL) accretion was
first discovered in 2--D numerical simulations by Matsuda \etal
(1987) for axisymmetric accretion and by Fryxell \& Taam (1988), Taam
\& Fryxell (1989) for flows including density or velocity gradients.
The shock surface oscillates from one side of the accretor to the
other (so called ``flip--flop" instability), leading to
high--amplitude, quasi--periodic variations of the mass accretion
rate.
This phenomenon was later confirmed by Matsuda \etal (1989,
1991, 1992), who showed that this process is more violent for small
accretor sizes, non absorbing accretors, and high Mach numbers
(see also Benensohn \etal 1997, Shima \etal 1998). 3--D
numerical simulations were performed by Ishii \etal (1993), Ruffert
(1996 and previous works for homogeneous media and 1997 for flows including
gradients), showing again quasi--periodic variations of the mass
accretion rate, although with a smaller amplitude (up to 30 per
cent), and with deformations of the shock surface only in the
immediate vicinity of the accretor.

Livio (1992) proposed a series of possible observational implications 
of the instability. In particular, it ought to occur in the accretion 
process of a neutron star orbiting in a dense wind in high mass X-ray 
binaries (HMXB)
(Taam \etal 1988, De Kool \& Anzer 1993). It was also applied to the 
supermassive black hole SgrA$^*$ at the galactic 
center (Ruffert and Melia 1994), and even to individual 
galaxies in the intracluster gas (Balsara \etal 1994). 

Accretion onto a point like Newtonian accretor moving supersonically in 
a uniform adiabatic
gas is of course a highly idealized problem. It presents the advantage 
of depending only on two dimensionless parameters, namely the adiabatic 
index $\gamma$ of the gas and its mach number $\M_\infty$ at infinity. 
Although this academic formulation seems simple, it gives rise to an 
instability for which no clear criterion is available yet. The extreme 
simplicity of this formulation leads us to expect simple laws describing 
the onset of instability, and in particular the distribution of 
timescales characterizing the instability. Numerical simulations have to 
include a third dimensionless parameter, namely the size of the accretor 
$r_*$ in units of the accretion radius ($\rA\equiv2GM/v_\infty^2$). Due 
to prohibitive computationnal cost, the smallest accretor size 
considered in 3--D was $r_*/\rA = 0.02$ (Ruffert 1996 and previous 
works). We would like to be 
able to extrapolate the results obtained with numerical simulations to 
smaller accretors, like a weakly magnetized neutron star or 
a black hole, moving at $v_\infty =1000$ km s$^{-1}$, for which 
$r_*/\rA \sim 10^{-5}$. 

Despite numerous numerical simulations, our 
under\-standing of the instability is still unsatisfactory. The 
instability of the accretion column for cold flows is well established
(Cowie 1977, Soker 1990, 1991), but does not directly apply to the case 
of hydrodynamic BHL flows. A tentative explanation for the flip--flop 
instability was proposed by Livio \etal (1991), who showed that a 
conical shock becomes unstable when its opening angle exceeds a 
critical value. However, recent numerical simulations 
(Ruffert 1994b, 1995) suggest that the origin of the instability 
in 3--D is to be found within the subsonic flow near the accretor 
rather than at the shock surface.
This lack of a definite physical explanation for the instability left 
open the question of whether this instability is influenced by numerical
artifacts or is a natural consequence of the physics involved.

A detailed understanding of the instability mechanism should enable us 
to predict how the instability is influenced by the effects of the
accretor size, density and velocity gradients in the upstream flow 
(Ruffert \& Anzer 1995, Ruffert 1997), radiative cooling and heating of 
the gas (Blondin \etal 1990, Taam \etal 1991) and relativistic effects 
near the accretor (Petrich \etal 1989, Font \& Ibanez 1998a,b), and be 
more confident in invoking this instability for the variability of 
accreting systems in astrophysics.

To suggest such a physical mechanism is the purpose of the present paper.

The local approach that we use is shortly reviewed in
Sect.~\ref{slocal}. A lower bound for the entropy gradient produced
by a shock is computed in Sect.~\ref{sectentr}.
We use a simplified formulation of the
Rayleigh--Taylor (Sect.~\ref{secRT}) and Kelvin--Helmholtz
(Sect.~\ref{sKH}) linear instabilities in order to estimate their 
influence on the stationary BHL flow described in Foglizzo \& Ruffert 
(1997, Paper~I). They are compared and discussed in Sect.~\ref{secdisc}.
The results of new subsonic 3-D simulations are interpreted
in the light of this analysis in Sect.~\ref{snew}.

\section{Local stability analysis\label{slocal}}

\subsection{Local linear growth rate integrated along a flow line}

According to the numerical simulations (e.g.~Ruffert 1995), the
instability of the BHL flow occurs only when a shock is present. The
shock is a source of entropy inhomogeneities and vorticity, and
therefore potentially a source for two well
known local instabilities: entropy gradients in a gravitational field
may lead to the Rayleigh--Taylor instability (hereafter \RTi), and
vorticity can induce the Kelvin--Helmholtz instability (hereafter
\KHi). Note that entropy gradients ${\vec \nabla} S$ and vorticity
${\vec w}$ in the BHL flow are closely related by Eq.~(10) in
Paper~I:
\begin{equation}
{\vec w}\times {\vec v}=T{\vec \nabla} S.\label{vortentro}
\end{equation}

Unlike Garlick (1979) and Petterson \etal (1980) who used a global
analysis to show the stability of the spherically symmetric accretion
flows, we adopt here a local approach to estimate the effect of the
\RTi and \KHi on the axisymmetric BHL flow. Although a global
perturbative analysis would in principle lead to conclusive
statements about the stability of the flow, it seems to be excessively
difficult for axisymmetric flows, where a boundary value problem must be
solved in two dimensions (radial and azimuthal) for perturbations of an
incompletely known stationary flow (Paper~I). The local approach has the
double advantage of being mathematically tractable and physically
understandable, although it might require some strong approximations.
Using the same notation as in Paper~I, flow lines are indexed by their
distance $\varpi$ to the symmetry axis at infinity. We evaluate the
typical local growth rate of each instability and integrate it
over the time available for amplification, \ie along a flow line $\varpi$
between the shock $r_{\rm sh}(\varpi)$ and the accretor surface
$r_\star(\varpi)$. We would like to check whether such
a mechanism can amplify perturbations up to non--linear
amplitudes before they are advected onto the surface of the accretor.
We consider the linear growth rate $|\growth_i({\vec r})|$ of the
instability ($i=$ RT or KH) as obtained from a normal mode approach
in an infinite medium of same entropy gradient and vorticity as at
the position ${\vec r}$ in the BHL flow. As stressed by Garlick (1979) 
in the case of spherical accretion, such a local approximation
is justified only if the distance over which the growth rate varies
$(\p\log\growth_i/\p r)^{-1}$ is longer than the distance
$(v/\growth_i)$ traveled during a growth time. The variation of the
growth rate is due to the convergence and acceleration of the flow,
typically on a scale $r$. The criterion can be stated quantitatively
as follows:
\begin{equation}
\left|{\growth_i({\vec r}) r\over {\vec v}}\right| \gg
\left|{\p \log\growth_i\over\p\log r}\right|\sim 1\;. \label{WKB}
\end{equation}
We estimate the quantity ${\cal A}_i(\varpi)$ defined as:
\begin{eqnarray}
{\cal A}_i(\varpi)&\equiv&\int_{t_{\rm sh}}^{t_{\star}}
|\growth_i({\vec r})| \d t= 
\int_{{\vec r}_{\rm sh}(\varpi)}^{{\vec r}_\star(\varpi)}
{|\growth_i({\vec r})|\over v({\vec r})} \d l\;, \label{EcalA}\\
\MA&\equiv&{\rm Max}\{ {\cal A}_i(\varpi), \varpi>0\},\label{defma}
\end{eqnarray}
where we defined the elementary displacement as $\d l\equiv v\d t$.
$\MA$ is a dimensionless number which depends on the three
dimensionless parameters of the problem, namely the Mach number
$\M_\infty\ge0$ of the flow at infinity, the adiabatic index
$5/3\ge\gamma\ge1$ and the accretor radius $r_\star\ge0$. We aim
at determining the maximum value of $\MA$ when these parameters are
varied.

The WKB approximation gives accurate results where its criterion 
(\ref{WKB}) is satisfied, \ie if $\MA\gg1$. A threshold $\MA^* $ exists 
above which the results are still significant, \eg within $10\%$. 
As shown by Bender \& Orszag (1978) in many illustrative examples of the WKB 
method, the WKB approximation often gives reliable results even when its 
criterion is marginally satisfied. Although this may lead us to expect
$\MA^*\sim 1$, the threshold $\MA^* $ does not always strictly equal 
unity and naturally depends on the problem considered. The exact 
determination of $\MA^*$ lies beyond the scope of this paper, and we shall
assume it is of the order of unity. The three following situations might be 
encountered:
\par (i) if $\MA\gg \MA^*$, we conclude that the criterion~(\ref{WKB})
is fulfilled, and that a strong linear instability is identified for the
corresponding set of parameters.
\par (ii) If $\MA>\MA^*$ is finite, we would conclude that a marginal linear 
instability is present, which may lead to a non--linear instability or 
saturation if the typical amplitude of the perturbations is larger than 
$\exp(-\MA)$. For example, $\MA\sim3$ would be enough to amplify to 
non--linear amplitudes initial perturbations of order $5\%$. 
\par (iii) If $\MA<\MA^*$, the criterion~(\ref{WKB}) is not fulfilled and 
our method does not allow us to reach a conclusive statement.

Note that none of the 3--D numerical simulations suggest a particularly
violent linear instability, since several accretion times $\rA/c_\infty$
are usually needed before the instability becomes visible 
(\eg Ruffert 1995). Case (i) is therefore a priori excluded in the range
of parameters covered by these simulations, \ie $\M_\infty \le 10$, 
$r_\star/\rA \le 0.02$.

\subsection{Local expansion in the vicinity of a point like accretor}

Our local approach makes the important ``continuity" assumption that the 
BHL flow on a Newtonian accretor of size $r_\star$ resembles the BHL 
flow on a point like Newtonian accretor when $r_\star\to0$. This allows 
us to make series expansions in the vicinity of the singularity $r=0$,
and check whether $\MA$ diverges when $r_\star\to 0$. Although the
limit $r\to0$  with Newtonian gravity is unrealistic
(see \eg Petrich \etal 1989, Font \& Ibanez 1998a,b for relativistic 
effects), it is useful in order to understand 
formally classical flows, before more sophisticated effects are added.

\section{Entropy distribution produced by the shock\label{sectentr}}

\subsection{Entropy gradient along the shock}

\begin{figure}
\psfig{file=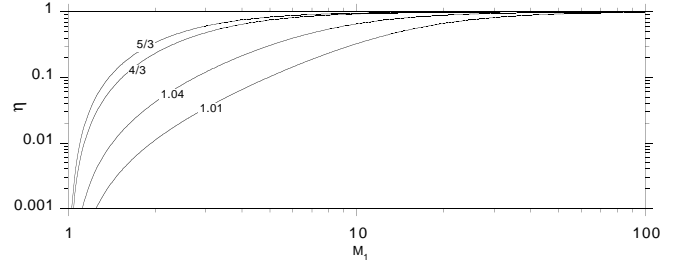,width=\columnwidth}
\caption[]{Coefficient $\eta<1$ entering the expression (\ref{ds2}) of the
entropy gradient created by the shock. The value of the adiabatic index
$\gamma$ is indicated on each curve. } \label{dS}
\end{figure}

For the sake of simplicity, in what follows we choose the same units 
as in Paper~I, such that the ratio 
of the mean molecular weight to the gas constant $\mu/{\cal R}=1$,
without loosing any generality.
The Rankine-Hugoniot conditions (\eg Landau \& Lifshitz 1987) 
imply that the entropy jump $\Delta S$ across an adiabatic shock is an
increasing function of the Mach number $\M_1$ associated to the velocity
component $v_{1\perp}$ ahead of and perpendicular to the shock:
\begin{equation}
\e^{\Delta S}=
\left\lbrack{2+(\gamma-1)\M_1^2
\over(\gamma+1)\M_1^2}\right\rbrack^{\gamma\over\gamma-1}
\left\lbrack
{2\gamma\M_1^2-\gamma+1\over\gamma+1}
\right\rbrack^{1\over\gamma-1}\;.
\label{entroRH}
\end{equation}
Let $v_{2\perp}$ be the velocity component 
perpendicular to and immediately after the shock.
We write the entropy gradient immediately after the shock as a function of 
$\M_1$ and its variations with respect to the curvilinear abscisse $L$ along 
the shock, using Eq.~(\ref{entroRH}):
\begin{eqnarray}
\nabla S&=& {2\eta\over\gamma-1}
{v_{2}\over v_{2\perp}}{\p \log \M_{1}\over\p L},
\label{ds2}\\
\eta&\equiv &{2\gamma(\gamma-1)(\M_1^2-1)^2\over
(2+(\gamma-1)\M_1^2)(2\gamma \M_1^2-(\gamma-1))} <1.
\label{defeta}
\end{eqnarray}
$\eta$ always converges to unity for large Mach numbers, when the kinetic 
energy of the gas exceeds its internal energy (\ie $\M_1^2\gg 2/(\gamma-1)$). 
The convergence is thus much slower for $\gamma$ close to 1 (Fig.~\ref{dS}). 

According to numerical simulations, the shock distance seems to vary
strongly from about 0.2 accretion radii for $\gamma=5/3$ to apparently 
zero for $\gamma=1.01$ ($r_\star=0.02$ in Ruffert 1996). It is not clear 
yet whether the shock would be detached for smaller accretors. 
Wolfson (1977) remarked that for $\gamma$ close to one, energy is 
soaked up by the internal degrees of freedom of the gas, therefore not 
contributing to support the shock through the kinetic pressure, thus 
favouring an attached shock. This leads us to consider successively the 
cases of attached and detached shock.

\subsection{Case of a shock attached to a point like accretor}

We deduce from Eqs.~(E1), (E2), (E4) in Paper~I that along a shock
attached to a point like accretor with an angle $\theta_{\rm sh}$, for 
$r\ll \rA$, the velocity scales like:
\begin{eqnarray}
v_{1\perp}&=&\left({2GM\over L}\right)^{1\over2}
\cos{\theta_{\rm sh}\over2}\sim{\gamma+1\over\gamma-1}v_{2\perp},
\label{v1pe}\\
v_{1\parallel}&=&\left({2GM\over L}\right)^{1\over2}
\sin{\theta_{\rm sh}\over2}=v_{2\parallel}.\label{v1pa}
\end{eqnarray}
Using Eq.~(\ref{ds2}), the entropy gradient immediately after the shock,
close to the point like accretor is:
\begin{equation}
\left|\nabla S\right|\sim{1\over\gamma-1}
\left\lbrack1+\left({\gamma+1\over\gamma-1}\right)^2
\tan^2{\theta_{\rm sh}\over2}\right\rbrack^{1\over2}{1\over L}.
\label{sattached}
\end{equation}
The entropy produced by the shock is a decreasing function of $L$
if the shock is attached. It depends on the Mach number only through the
shock opening angle $\theta_{\rm sh}$.

\subsection{Case of a detached axisymmetric shock\label{secdetsh}}

Let $\rs(\theta)$ be the shape of the axisymmetric shock
surface, using polar coordinates $(r,\theta)$ centred on the accretor.
\begin{figure}
\psfig{file=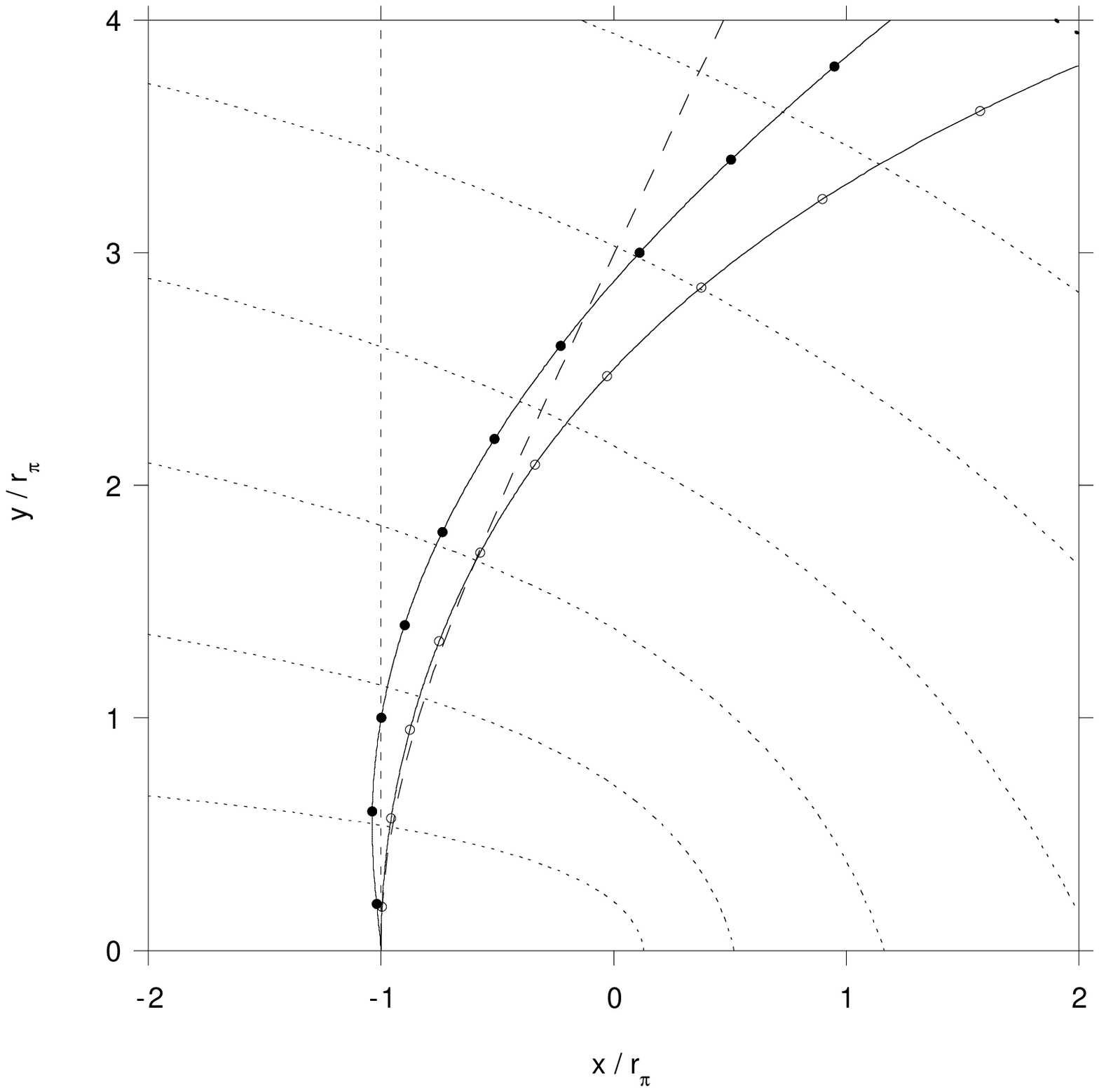,width=\columnwidth}
\psfig{file=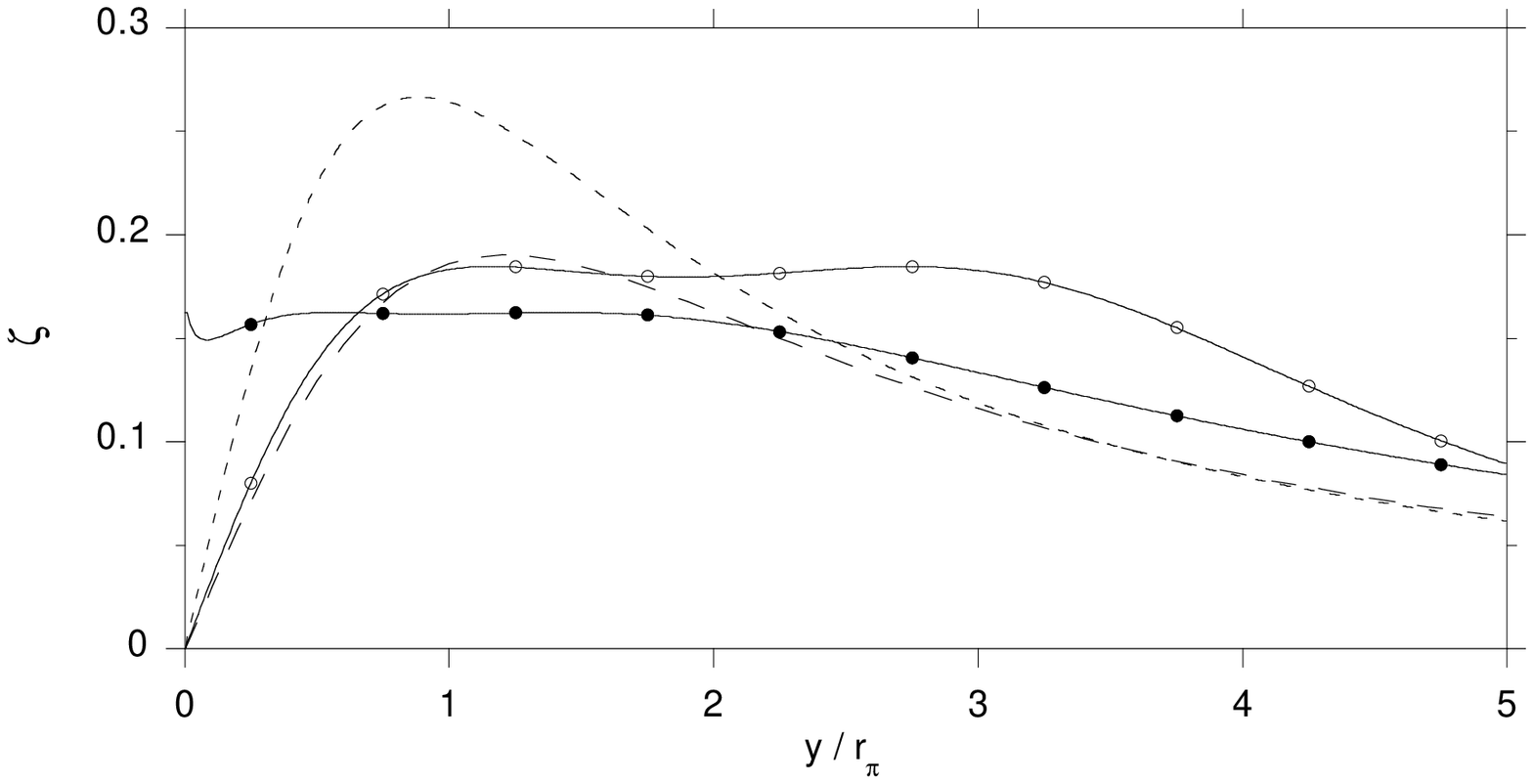,width=\columnwidth}
\caption[]{The coordinates of the accretor are $(0,0)$ on the upper plot.
The hyperbolic trajectories of the gas emanating from $x<0$ are 
represented by dotted lines. Four particular shock shapes are plotted for 
$r_\pi=0.1$ accretion radii. The short dash line is orthogonal to the 
symmmetry axis, and the long dash curve is orthogonal to the supersonic 
flow lines. The curves with circles are the optimal shapes leading to the 
smallest entropy gradients corresponding to polynomial shapes of the variable 
$y^2$ (empty circles) and $y^{1/2}$ (filled circles). The value of $\zeta$ 
along these four curves is displayed on the bottom plot.}
\label{figmini}
\end{figure}
Let $r_\pi\equiv \rs(\pi)$ be the distance of the shock from the accretor, 
along the symmetry axis. Eq.~(\ref{entroRH}) indicates
that when $\M_\infty\gg1$, the entropy jump along the symmetry axis is:
\begin{equation}
\Delta S(r_\pi) \sim{2\over\gamma-1}\log \M_1
\ge {2\over\gamma-1}\log \M_\infty\;.\label{deltamax}
\end{equation}
Far from the accretor, the shock surface approaches the
Mach cone of semi-angle $\theta_s$ defined by
$\sin \theta_s\equiv 1/\M_\infty$ ($\M_1\sim 1$). 
The entropy jump therefore decreases from
$\Delta S(r_\pi)$ ahead of the accretor to zero far from it.
Since the entropy gradient immediately after the shock surface 
vanishes both on the symmetry axis and far from the accretor, the  
maximum entropy gradient $\left|\nabla S\right|_{\rm max}$ is reached
on a circle $\rs(\theta_{\rm max})$ corresponding to an intermediate azimuthal
angle $\theta_{\rm max}$. \\
Let us denote by $\alpha$ the angle between the flow line and the vector 
perpendicular to the shock surface, before the shock, so that 
$v_{1\perp}= v_1 \cos\alpha$. Defining the dimensionless function $\zeta$ 
along the shock surface $\rs(\theta)$ as
\begin{equation}
\zeta(\rs(\theta))\equiv -{r_\pi\over\cos\alpha}{\p \log \M_{1}\over\p L}
\le \zeta_{\rm max},
\label{defzeta}
\end{equation}
Eq.~(\ref{ds2}) is rewritten as follows:
\begin{eqnarray}
\nabla S&=& -{2\eta\over\gamma-1}\left\lbrack 1+
\left({v_{1\perp}^2\over v_{2\perp}^2}-1\right)\sin^2\alpha
\right\rbrack^{1\over2}
{\zeta\over r_\pi},\label{ds22}\\
1&<&{v_{1\perp}\over v_{2\perp}}
={(\gamma+1)\M_1^2\over2+(\gamma-1)\M_1^2}
<{\gamma+1\over\gamma-1}.\label{RHv}
\end{eqnarray}
The Rankine-Hugoniot condition (Eq.~\ref{RHv}) and Eq.~(\ref{ds22}) 
provide us with both a lower and an upper 
bound for the maximum entropy gradient produced by a strong shock 
($\eta\sim 1$ for $\M_\infty\gg1$):
\begin{equation}
{2(\gamma+1)\over(\gamma-1)^2}{\zeta_{\rm max}\over r_\pi}
\ge \left|\nabla S\right|_{\rm max}\ge 
{2\over\gamma-1}{\zeta_{\rm max}\over r_\pi}.
\label{encadre}
\end{equation}
The value of the maximum $\zeta_{\rm max}$ of the function $\zeta$ 
depends only on the properties of the supersonic flow before the shock,
and on the shape of the shock surface $\rs(\theta)$.
The function $\zeta^*(r_\pi)$ is defined as the minimum value of
$\zeta_{\rm max}$ for all possible continuous mathematical curves 
$\rs(\theta)$ satisfying $\rs(\pi)=r_\pi$ and $\d \rs/\d \theta(\pi)=0$:
\begin{equation}
\zeta^*(r_\pi)\equiv {\rm Min}\;\left\lbrace\zeta_{\rm max},\;
{\rm any \;curve}\;r_{\rm sh}(\theta)\;,\rs(\pi)=r_\pi
\right\rbrace.\label{defzetastar}
\end{equation}
Thus we obtain a lower bound for the 
maximum entropy gradient $|\nabla S |_{\rm max}$ produced by 
a detached shock standing at the distance $r_\pi$ from the accretor:
\begin{equation}
\left|\nabla S\right|_{\rm max}\ge {2\over\gamma-1}
{\zeta^*\over r_\pi}.\label{entropapprox}
\end{equation}
\begin{figure}
\psfig{file=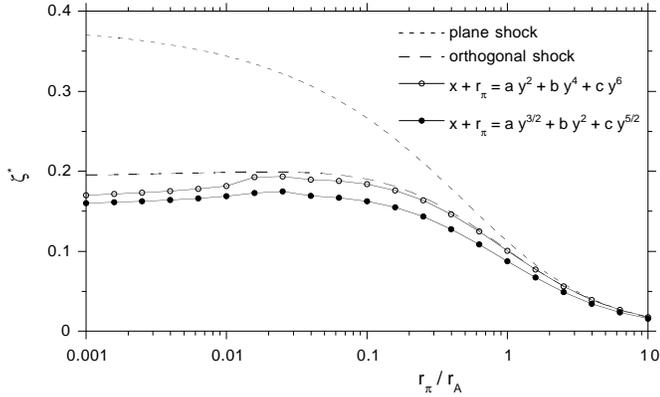,width=\columnwidth}
\caption[]{Coefficient $\zeta^*$ characterizing the minimum value of the 
maximum entropy gradient immediately after an axisymmetric shock, depending 
on its distance $r_\pi$ to the accretor. The shock shapes are the same as in 
Fig.~\ref{figmini}. }	
\label{figzeta}
\end{figure}
In the framework of the approximation of the supersonic trajectories by 
hyperbolae, for $\M_\infty\gg1$, $\zeta^*$ depends only on the distance 
$r_\pi$ of the shock. We have computed numerically the function 
$\zeta^*(r_\pi)$ using a polynomial approximation of the shock shape
and a downhill simplex method for the minimization (Press \etal 1992). 
Powell's method was also used with comparable results. Satisfactory results 
were obtained with a polynomial $x(y)$ of order 3 in $y^2$. The overall 
minimum seems to be reached by the singular curve scaling like 
$x+r_\pi\sim y^{3/2}$ near the symmetry axis, which can be approached by a 
series of regular polynomials (Figs~\ref{figmini} and \ref{figzeta}). 
For comparison, a plane shock orthogonal to the symmetry axis
produces typically twice as much entropy gradients than the minimum value
($\zeta_{\rm max}/\zeta^*\sim 2$).  
More realistic is the curve orthogonal to the 
supersonic flow lines ($\alpha\equiv0$), which produces entropy 
gradients about $20\%$ stronger than the absolute minimal value.
According to Fig.~\ref{figzeta}, the coefficient $\zeta^*> 0.1$ for 
realistic shock distances, \ie $r_\pi < \rA$. The maximum entropy 
gradient along the shock stands near $L\sim r_\pi$ (Fig.~\ref{figmini}). 

Fig.~\ref{figmini} shows that the mathematical curves approaching the 
minimum value of the entropy gradient after the shock are not very different 
from the physical shock shapes observed in numerical simulations 
(\eg Fig.~\ref{vortflow}). The lower bound $\zeta^*$ might therefore be a 
good approximation of the realistic value of $\zeta_{\rm max}$, within a factor
two.

\subsection{Entropy gradient in the subsonic flow between a detached 
shock and the accretor}

If the boundary condition on the surface of the accretor allows
a high enough mass accretion rate, the maximum entropy gradient 
immediately after the shock corresponds to a flow line $\vmax$
converging to the accretor.
With the entropy remaining constant along each flow line after the shock, 
the gradient of entropy across the flow is simply the gradient immediately 
after the shock surface modified by a geometrical factor. If the 
convergence towards the accretor were along straight flow lines, the 
entropy gradient would simply increase like $1/r$. However, flow lines 
are not radial, and we can take this into account by including 
a geometrical factor $\delta({\vec r})$, so that the distance between 
two neighbouring flow lines converging towards the accretor scales 
like $r/\delta({\vec r})$:
\begin{equation}
\delta({\vec r})\equiv{r(\theta)\over \rs} 
{\nabla S({\vec r})\over(\nabla S)_{\rm sh}}
\;,
\label{defdelta}
\end{equation}
where $r(\theta)$ is the shape of the flow line $\vmax$, and
$(\nabla S)_{\rm sh}$ is the entropy gradient on this flow line
immediately after the shock, at a distance $\rs$, so that 
$\delta({\vec r}_{\rm sh})=1$. 
We made a distinction in Sect.~4 of Paper~I between the directions 
of regular and singular accretion on a point like accretor. The 
distance between adjacent accreted flow lines decreases like $\sim r$ 
when $r\to0$ in a direction of regular
accretion ($\lim_{r\to0}\delta(r)$ is finite), whereas it decreases much
faster in a direction of singular accretion ($\lim_{r\to0}\delta(r)$
diverges). Since we proved in Paper~I that accretion for a gas with
$\gamma=5/3$ is always regular, we know that the instability of the
BHL flow does not rely on the presence of directions of singular 
accretion. We can therefore assume that $\lim_{r\to0}\delta$ is finite 
in our analysis of the instability.

We obtain from Eqs.~(\ref{encadre}) and (\ref{defdelta}) 
upper and lower bounds for the entropy gradient between the detached
shock and the accretor, along the flow line $\vmax$, 
at high Mach number:
\begin{equation}
{2(\gamma+1)\zeta_{\rm max}\over (\gamma-1)^2}
{\rs\over r_\pi}{\delta\over r}\ge
\left|\nabla S\right|_{\rm max}
\ge {2\zeta_{\rm max}\over \gamma-1}{\rs\over r_\pi}{\delta\over r}
\;.\label{entropapprox2}
\end{equation}
Note that the ratio $\rs/ r_\pi$ is of the order of $2^{1\over2}$ in 
Fig.~\ref{figmini}.

\section{Rayleigh--Taylor instability\label{secRT}}

\subsection{A simplified formulation of the RT instability}

Let us consider a stratified gas in a gravitation field.
An entropy gradient can act either in a stabilizing or
destabilizing manner depending on whether it does or does not
contribute to support the gas against gravity. The vertical
gravity field ${\vec g}$
and pressure forces at equilibrium are related as follows:
\begin{equation}
{\vec g}={1\over \rho_0}{\vec\nabla}P_0\;.\label{equili}
\end{equation}
The Brunt--V\"ais\"al\"a frequency $N$ is the frequency of oscillations of
the stratified gas, in the limit of small wavelengths perpendicular to the
gravity field (see Appendix~\ref{aprtsh}). 
$\growth_{\rm RT}\equiv (-N^2)^{1\over2}$
is the local growth rate of the \RTi when the entropy decreases upwards:
\begin{equation}
\growth_{\rm RT}^2\equiv -N^2\equiv{\gamma-1\over \gamma}
{\vec g}\cdot{\vec\nabla} S \;.\label{defsrt}
\end{equation}
If the flow is sheared with height, perturbations
with a short horizontal wavelength in the direction perpendicular to
the flow velocity grow at the same rate (see Appendix~\ref{aprtsh}).

\subsection{Effective gravity in the comoving frame\label{sectg}}

\begin{figure}
\psfig{file=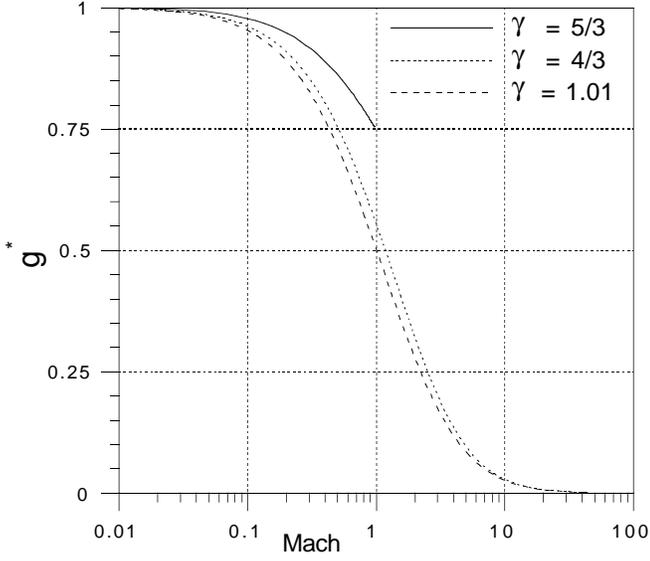,width=\columnwidth}
\caption[]{Effective gravity, in $GM/r^2$ units, in the case of spherical
accretion for various values of $\gamma$.}
\label{graveff}
\end{figure}
From the principle of equivalence, there would be no \RTi in a gas falling
freely in a gravitational potential, because the effective gravity would
then be zero.  So in a reference frame falling with the gas, the effective 
gravity $g_{\rm eff}$ driving the \RTi is opposite to the pressure force:
\begin{equation}
{\vec g}_{\rm eff}\sim {1\over\rho}{\vec\nabla}P\;.\label{defgeff}
\end{equation}
According to Sect.~4.6 in Paper~I, the pressure 
close to the accretor is spherically symmetric 
to first order for $\gamma=5/3$. This leads us to neglect the negative 
contribution of the azimuthal pressure force to the scalar product 
${\vec \nabla} P\cdot{\vec\nabla} S$ in Eq.~(\ref{defsrt}).
The radial pressure support decreases from the subsonic region to the
supersonic region. The effective gravity, calculated analytically in the
case of spherical accretion, is displayed in Fig.~\ref{graveff} for various
adiabatic indices. In the subsonic region, the effective gravity is more
than $50\%$ of the gravity of the accretor for any value $1<\gamma<5/3$.
The best pressure support is reached, of course, for $\gamma=5/3$,
for which the effective gravity in the subsonic region is at least
$75\%$ of the gravity. We shall assume that the effective gravity for an
axisymmetric flow is comparable to the effective gravity in the
spherical case, thus constraining the dimensionless gravity parameter
$g^*\in [0.5,1]$ in the subsonic region of the flow:
\begin{equation}
g^*\equiv g_{\rm eff} {r^2\over GM}.\label{defg*}
\end{equation}

\subsection{\RTi efficiency in the BHL flow\label{secRTeff}}

\subsubsection{General expression of the RT efficiency}

According to Eq.~(\ref{defsrt}), the local growth rate of nonaxisymmetric 
perturbations with a short wavelength perpendicular to
both the flow lines and the radial direction (similar to the case
$k_x\to\infty$, $k_y=0$ in Appendix~\ref{aprtsh}) is approximated as follows:
\begin{equation}
\growth_{\rm RT}^2= {\gamma-1\over \gamma}
g^* {GM\over r^2}|\nabla S|\sin\beta\;,
\label{sRT}
\end{equation}
where $\beta$ is the angle between the flow line and the radial
direction:
\begin{equation}
\tan\beta\equiv {v_\theta\over v_r}.\label{defbeta}
\end{equation}
The perpendicular wavelength of a non axisymmetric perturbation decreases 
geometrically as the flow is advected, so that the growth rate of the 
\RTi stays maximum. 

Defining the free fall velocity $\vff$ as
\begin{equation}
\vff\equiv \left({2GM\over r}\right)^{1\over2},
\label{vff}
\end{equation}
the integrated efficiency of the \RTi defined by Eq.~(\ref{EcalA}) 
follows from Eqs.~(\ref{sRT}) and (\ref{vff}) :
\begin{eqnarray}
{\cal A}_{\rm RT}(\varpi)&\equiv&
\left({\gamma-1\over 2\gamma}\right)^{1\over2}\times\nonumber\\
&&\int_{{\vec r}_{\rm sh}(\varpi)}^{{\vec r}_\beta(\varpi)}
g^{*{1\over2}}{\vff\over v}
\left(r|\nabla S|\right)^{1\over2}
\sin^{1\over2}\beta {\d l\over r}\;. \label{ERTgeneral}
\end{eqnarray}
With the estimates of the entropy gradient and effective gravity of
Sects.~\ref{sectentr}
and \ref{sectg}, we are now able to evaluate Eq.~(\ref{ERTgeneral}) for
the different
topologies discussed in Paper~I. 
Since the entropy along the shock decreases away from the symmetry
axis, the stratification is potentially linearly unstable only in the
region where $\beta>0$. The conservation of angular momentum implies
that $\beta>0$ in the supersonic flow before the shock. Since $\beta$ 
is likely to increase across the shock, we conclude that the stratification 
is locally unstable immediately after the shock surface.
We denote by $r_{\rm \beta}(\theta)\ge 0$ the surface where the velocity is 
radial, thus delimiting a region of unstable stratification.

\subsubsection{Attached shock\label{secattached}}

Using Eqs.~(\ref{sattached}), (\ref{sRT}) close to accretor ($\eta\sim 1$),
and the relation $\sin\beta=v_{2\perp}/v_2$, the local growth rate is:
\begin{equation}
\growth_{\rm RT}=
\left({g^*\over2\gamma}\right)^{1\over2}{\vff\over r}.
\end{equation}
This growth rate must be integrated along a flow line $\varpi$ between the
azimuthal angles $\theta_{\rm sh}$ and $\theta_{\rm so}$ corresponding to 
the shock and the sonic surfaces respectively. The length of this path of
integration is of the order of $ r (\theta_{\rm sh}-\theta_{\rm so}$). 
Estimating the velocity after the shock from Eqs.~(\ref{v1pe}) and 
(\ref{v1pa}) gives:
\begin{eqnarray}
{\cal A}_{\rm RT}(\varpi)&\equiv&
\int^{\theta_{\rm so}}_{\theta_{\rm sh}}
{\growth_{\rm RT}\over v}\d l,\\
&\sim&{\gamma+1\over2\gamma}
\left({g^*\over 2}\right)^{1\over2}
{\theta_{\rm sh}-\theta_{\rm so}\over
\left\lbrack
\sin^2{\theta_{\rm sh}\over2}+{(\gamma-1)^2\over4\gamma}
\right\rbrack^{1\over2}}<2^{1\over2}.\label{attached}
\end{eqnarray}
The efficiency depends strongly on the azimuthal size of the subsonic region
reaching the accretor ($\theta_{\rm sh}-\theta_{\rm so}$). This parameter is
unfortunately unknown, and we only obtain an upper bound on the efficiency of 
the \RTi: the timescale of the \RTi instability is at best comparable to the 
advection timescale for a shock attached to the accretor, however small the 
accretor might be. 

\subsubsection{Region of supersonic accretion near a point like accretor 
with a detached shock, $\gamma<5/3$}

The sign of $\beta$ in the supersonic region might simply preclude the
instability ($\beta<0$ for $\gamma=5/3$). Let us show that even if the 
flow lines were bent in the unstable direction 
($\beta>0$ for $\gamma\sim 1$), the \RTi would become negligible when the
gas approaches a point like accretor. For this we wish to check that the
divergence of the entropy gradient when $r\to 0$ is not fast enough to 
make the \RTi growth time shorter than the free fall time.
Using Eq.~(\ref{entropapprox2}) and Eq.~(\ref{sRT}), the radial dependence of 
the \RTi growth rate in the supersonic region scales like:
\begin{eqnarray}
\growth_{\rm RT}(r)&=&{\cal O}\left(
g^{*{1\over2}}{\beta^{1\over2}\over r^{3\over2}}\right)\\
&\ll&{\cal O}\left(r^{-{3\gamma+1\over4}}\right),
\end{eqnarray}
where we have used Eq.~(56) of Paper~I for an upper bound of $\beta(r\to0)$
($\p\log\beta/\p\log r\ge (5-3\gamma)/2$), and the decrease of the 
effective gravity.
Using the free fall approximation of the velocity, the contribution of the
region surrounding a point like accretor to the efficiency of the \RTi 
scales like:
\begin{equation}
\int^{{\vec r}}_0
{\growth_{\rm RT}\over v}{\d r\over \cos\beta}\ll{\cal O}
\left(r^{5-3\gamma\over2}\right)\;.
\label{ARTff}
\end{equation}
From the convergence of the integral when $r\to0$ we deduce that the
local growth time is much longer than the free fall time, and the \RTi can
be neglected there if $\gamma<5/3$. 

\subsubsection{Region of subsonic accretion along the maximum entropy
gradient\label{secsubacc}}

Since the \RTi is driven by the entropy gradient, it is natural to evaluate
a lower bound for its efficiency along the flow line $ \vmax $ associated 
with the maximum entropy gradient, at high Mach number ($\eta\sim1$), using Eqs.~(\ref{entropapprox2}) and (\ref{ERTgeneral}):
\begin{eqnarray}
{\hat{\cal A}}_{\rm RT} &\ge& {\rm Min}\;{\cal A}_{\rm RT},
\label{ARTmini}\\
{\rm Min}\;{\cal A}_{\rm RT}&\equiv &
\left({\zeta_{\rm max}\over\gamma}{\rs\over r_\pi}\right)^{1\over2}
\int_{{\vec r}_{\rm sh} }^{{\vec r}_\beta }
{\vff\over v}(\delta g^*\sin\beta)^{1\over2}
{\d l\over r}\;.
\label{ART}
\end{eqnarray}
The Bernoulli equation can be written in terms of the free fall velocity,
and approximated for $\M_\infty^2\gg 2/(\gamma-1)$, inside the sonic radius:
\begin{eqnarray}
v^2\left\lbrack 1+{2\over(\gamma-1)\M^2}\right\rbrack&=&
\vff^2 + v_\infty^2\left\lbrack 1+{2\over(\gamma-1)\M_\infty^2}\right\rbrack
\label{Bernoulli},\\
&\sim& \vff^2\label{bernapprox},
\end{eqnarray}
where we have neglected $v_\infty$ compared to $\vff$ between the shock and 
the accretor, since the shock distance is typically shorter than the accretion 
radius ($\vff(\rA)=v_\infty$). The ratio $\vff/v$ decreases in the 
subsonic region between the shock ($\rs$) and the sonic point 
($r_{\rm sonic}$). Applying Eq.~(\ref{bernapprox}) at the sonic point 
($\M=1$) and at the shock ($\M\sim\M_{\rm sh}$), we obtain the following range:
\begin{equation}
\left({\gamma+1\over\gamma-1}\right)^{1\over2}
\le {\vff\over v}\le {\gamma+1\over\gamma-1}.\label{range}
\end{equation}
If $\gamma<5/3$, the flow line $\vmax$ reaches the accretor in the
supersonic region, and we know from the preceding section that the 
contribution of this region to the integral is negligible.

Let us examine each of the terms of Eq.~(\ref{ART}) for $\gamma=5/3$, which
is supposed to be the most unstable case according to numerical simulations.
\par(i) $0.75\le g^*\le 1$ according to Fig.~\ref{graveff}, and thus we
estimate $g^*\sim 0.9$,
\par(ii) the geometrical factor $\delta$ is finite since accretion with 
$\gamma=5/3$ is always regular (Paper~I), and is assumed to be
of the order of unity,
\par(iii) $2\le v_{\rm ff}/v\le 4 $ according to Eq.~(\ref{range}), and thus 
we estimate $v_{\rm ff}/v\sim 3 $.

Because these three contributions to the integral are finite, we replace 
each of them by their mean value and approximate Eq.~(\ref{ART}) as follows:
\begin{equation}
{\rm Min}\;{\cal A}_{\rm RT}\sim 1.0\;
\left({\zeta_{\rm max}\over 0.15}\delta
{\rs\over 2^{1\over2}r_\pi}\right)^{1\over2}
\int_{{\vec r}_{\rm sh} }^{{\vec r}_\beta }
\sin^{1\over2}\beta{\d l\over r}\;.
\label{ART2}
\end{equation}
We showed in Paper~I that $\beta<0$ along the sonic surface for 
$\gamma=5/3$, thus $r_\beta(\varpi)>0$ can be estimated as a fraction of the
shock distance $r_\pi$. This precludes the possibility
that the integral in Eq.~(\ref{ART2}) might diverge when the size of the 
accretor is decreased. This further suggest that the efficiency of the \RTi
instability should not increase much when the accretor size is much 
smaller than the shock distance ($r_\star\ll r_\pi$).
In order to estimate this efficiency, let us remark from Fig.~\ref{vortflow}
that $\sin\beta$ is close to unity immediately after the shock (see also 
numerical simulations as in Fig.\ref{figmini}). Since we expect that 
$\beta>0$ over a sizable fraction of the shock distance,
this integral is likely to be of order unity. According to the typical
values of $\zeta_{\rm max}$ deduced from Fig.~\ref{figzeta}, we conclude 
from Eq.~(\ref{ART2}) that Min ${\cal A}_{\rm RT}$ 
is of the order of unity, hardly more. Since the lower and upper bounds in 
Eq.~(\ref{entropapprox2}) differ by a factor $(\gamma+1)/(\gamma-1)$, and since 
the \RTi growth rate scales like the square root of the entropy gradient, we 
finally estimate for $\gamma=5/3$:
\begin{equation}
1.0\la{\hat{\cal A}}_{\rm RT}\la2.0\; .\label{ARTfinal}
\end{equation}  
Thus the integrated efficiency of the linear \RTi does not diverge, even in 
the case of a point like accretor moving at high Mach number in a gas with
$\gamma=5/3$. Although small, this efficiency of order unity is not 
negligible.

\section{Kelvin--Helmholtz instability\label{sKH}}

\subsection{A simplified formulation of the KH instability}

\begin{figure}
\psfig{file=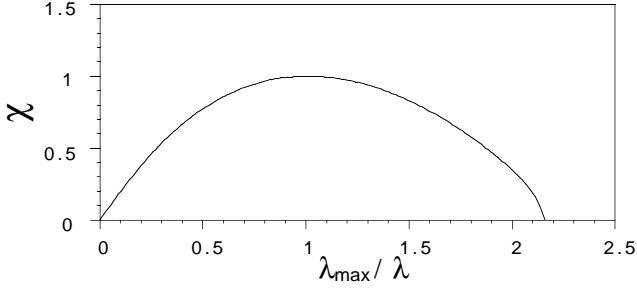,width=\columnwidth}
\caption[]{Typical dependence of the Kelvin-Helmholtz instability on
the wavelength $\lambda$ of the perturbation, described by the function
$\chi$ (full line). The dotted line shows the slope at long wavelengths.}
\label{chi}
\end{figure}

The maximum linear growth rate of the \KHi in an inviscid fluid is
comparable to the maximal vorticity of the flow in equilibrium. From
the studies of the instability with various velocity profiles (see an
overview in Drazin and Reid, 1981), one can extrapolate the following
growth rate $\growth_{\rm KH}$ and optimal wavelength $\lambda_{\rm
max}$ of the instability corresponding to a vorticity profile with a
maximum $|w_{\rm max}|$ and a gradient width $\grwdth$:
\begin{eqnarray}
|\growth_{\rm KH}|& = & \pttwo \;|w_{\rm max}|\; \chi({\lambda_{\rm
max}\over\lambda}),
\quad {\rm with}\,\, \alpha \approx 0.2, \label{ErateKH}\\
\lambda_{\rm max}& = & \seven\; \grwdth\;, \quad\quad \quad\quad
\quad\quad \quad \, {\rm with}\,\, \seven \approx 7,
\label{EwaveKH}
\end{eqnarray}
where $\chi(x)\le\chi(1)=1$ is a function of the wavelength $\lambda$
of the perturbation. Its typical shape, obtained by a numerical solution
of the Orr-Sommerfeld equation, is plotted in Fig.~\ref{chi}.

The effect of compressibility on the linear instability can be
studied by solving the linearized equations for a sheared plane flow
(see Appendix~\ref{akhc}), with the following velocity profile:
\begin{eqnarray}
v(z)&\equiv &{v_0\over 2}\tanh {2z\over \grwdth},\label{profil}\\
w_{\rm max}&=&{v_0\over \grwdth}.\label{profil2}
\end{eqnarray}
Neither the value of $\gamma$, within the range $[1,5/3]$, nor the presence 
of an entropy gradient across the flow influences the growth rate and
the wavelength of the most unstable mode by more than a few percent.

\begin{figure}
\psfig{file=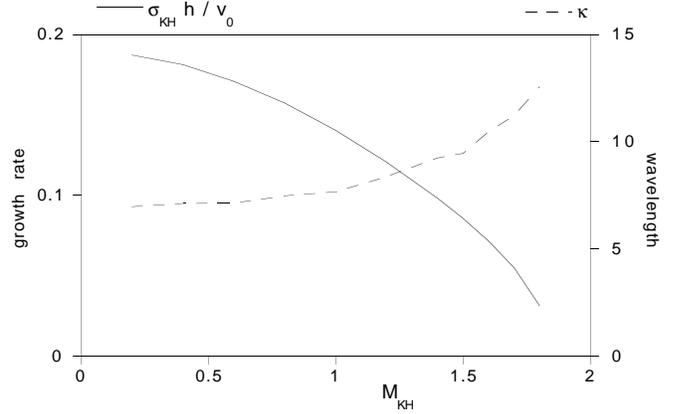,width=\columnwidth}
\caption[]{Maximum growth rate of the linear \KHi and
associated wavelength, as a function of the Mach number 
$\M_{\rm KH}=v_0/c_0$ for a uniform sound velocity $c_0$, and a
velocity profile given by Eq.~(\ref{profil}). 
The growth rate $\growth_{\rm KH}$ is displayed in units of the vorticity 
maximum $w_{\rm max}=v_0/\grwdth$ (solid line). The dimensionless parameter 
$\seven$ measures the optimal wavelength $\lambda_{\rm max}$ in units of 
the width of the vorticity peak $\grwdth$ (dashed line). The rigid 
boundaries are at $\pm 5\grwdth$. Within the accuracy of our numerical 
relaxation method, these curves are independent of the adiabatic index 
in the range of interest $1\le\gamma\le5/3$.}    
\label{figmach}
\end{figure}

A determining quantity for the \KHi in
compressible flows is the relative Mach number $\M_{\rm KH}$ measured
in the frame comoving with the flow line of maximum vorticity: 
\begin{equation}
\M_{\rm KH}\equiv {|v(\grwdth/2)-v(-\grwdth/2)|\over c_0} \sim
{\grwdth w_{\rm max}\over c_0}\;, \label{critcomp}
\end{equation}
with $c_0$ being the sound speed on the line of maximum vorticity
($z=0$), and $v(z)$ the $z$-dependent velocity of the flow in the
$y$--direction. The growth rate decreases by a factor of $2$ between the
very subsonic flow (which mimics the incompressible case) and
$\M_{\rm KH}=1.3$, as shown on Fig.~\ref{figmach}. So the difference of 
velocities within the vorticity peak must be less than the sound speed 
(\ie $\M_{\rm KH}<1$) in order to allow pressure forces to act in the \KHi 
mechanism as efficiently as in the incompressible limit. Note that the 
limiting values of $\growth_{\rm KH}$ and $\seven$ for $\M_{\rm KH}\to 0$ in 
Fig.~\ref{figmach} agree with the values of $\pttwo$ and $\seven$ stated in 
Eqs.~(\ref{ErateKH}) and (\ref{EwaveKH}).

\subsection{Application to the BHL flow with a detached shock
\label{secKHeff}}

According to Eq.~(\ref{vortentro}), the same symmetry arguments used 
for the entropy gradients in Sect.~\ref{sectentr} apply to the vorticity. 
A surface of steepest increase of the vorticity therefore connects the 
shock to the accretor, where the vorticity diverges for $r\to 0$ 
(Eqs.~49--59 in Paper~I).
\begin{figure}
\psfig{file=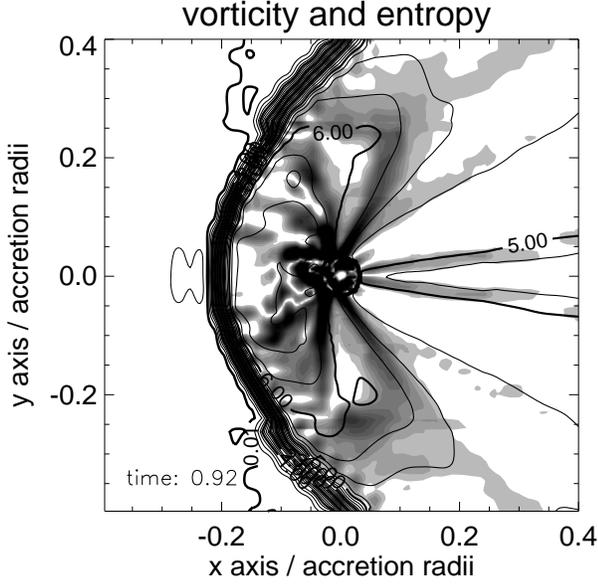,width=\columnwidth}
\caption[]{Lines of constant entropy superimposed on a greyscale map 
of the vorticity in a numerical simulation of BHL with $\gamma=5/3$ 
and $\M_\infty=10$ (model FS in Ruffert 1994b). 
In a good approximation, the line of steepest 
increase of the entropy gradient coincide with the line of steepest 
increase of the vorticity, and corresponds to a line of flow.}       
\label{vortflow}
\end{figure}
If this surface of maximum vorticity is not
too different from a flow surface, the \KHi will develop most
efficiently along these particular flow lines. Numerical
simulations, illustrated by Fig.~\ref{vortflow}, together with
Eq.~(\ref{vortentro}), suggest that the surface of steepest increase of
the entropy gradient and the surface of steepest increase of the vorticity
coincide with the flow line denoted by $\vmax$, such that $\theta_0(\vmax)$
is a direction of maximum entropy gradient and vorticity.

We use Eqs.~(\ref{vortentro}) and (\ref{entropapprox2}) along these flow 
lines to write the vorticity as follows:
\begin{eqnarray}
w_{\rm max}&=&{c^2\over \gamma v}\nabla S\;,\label{vormax}\\
&\ge&{2\zeta_{\rm max}\over \gamma (\gamma-1)}
{\delta\over \M^2}{\rs\over r_\pi}{v\over r}\;.
\label{Ewmax}
\end{eqnarray}
$\delta$ is related to the width $\grwdth(r)$ of the vorticity peak 
($\grwdth_{\rm sh}\equiv \grwdth(\rs)$), 
and thus to the optimal wavelength of the \KHi, through 
Eqs.~(\ref{defdelta}) and (\ref{EwaveKH}):
\begin{equation}
\delta({\vec r})={\grwdth_{\rm sh}\over\grwdth(r)} \left({r\over \rs}\right)
\;.  \label{delth}
\end{equation}
Although the vorticity defined in Eq.~(\ref{Ewmax}) may become arbitrarily
large when $\gamma$ is close to unity, the time-scale of the \KHi is limited 
by compressibility effects. Let us estimate the Mach number 
$\M_{\rm KH}$ defined in Eq.~(\ref{critcomp}), using  Eqs.~(\ref{Ewmax}) and
(\ref{delth}):
\begin{equation}
\M_{\rm KH}\sim {2\zeta_{\rm max}\over\gamma(\gamma-1)}
{\grwdth_{\rm sh}\over r_\pi}{1\over\M}, \label{limcomp}
\end{equation}
Paradoxically, Eq.~(\ref{limcomp}) indicates that compressibility effects
are stronger in subsonic regions. We introduce in Eq.~(\ref{limcomp}) the 
minimum value $\M_{\rm sh}$ of the Mach number after the shock, defined by the Rankine--Hugoniot jump conditions:
\begin{eqnarray}
\M_{\rm sh}&\equiv &\left({\gamma-1\over 2\gamma}\right)^{1\over2},
\label{mach2}\\
\M_{\rm KH}&\sim &{0.4\over\gamma^{1\over2}(\gamma-1)^{3\over2}}
{\zeta_{\rm max}\over 0.15}\;
{\grwdth_{\rm sh}\over r_\pi}\;{\M_{\rm sh}\over\M},\label{compeff}\\
&\sim &0.6\;{\zeta_{\rm max}\over 0.15}\;
{\grwdth_{\rm sh}\over r_\pi}\;{\M_{\rm sh}\over\M},\label{mkh53}
\end{eqnarray}
where Eq.~(\ref{mkh53}) assumes $\gamma=5/3$. Noting that the Mach number 
increases from the shock to the sonic surface ($\M>\M_{\rm sh}$), and 
that $\grwdth_{\rm sh}/r_\pi$ is of the order of unity, we conclude that the 
effect of compressibility on the \KHi can be neglected for $\gamma=5/3$.
From Eq.~(\ref{ErateKH}) and (\ref{Ewmax}), we can write the minimum \KHi 
growth rate, at high mach number ($\M_\infty\gg1,\eta\sim 1$), as follows:
\begin{equation}
|\growth_{\rm KH}|\ge {2\pttwo\zeta_{\rm max}\delta \over \gamma
(\gamma-1)}\;{ \chi\over \M^2}\;{\rs\over r_\pi}\;{v\over r}\;,
\label{wKH}
\end{equation}
Let us estimate the minimum efficiency Min ${\cal A}_{\rm KH}$ of the 
\KHi mechanism along the flow line $\vmax$:
\begin{equation}
{\rm Min}\;{\cal A}_{\rm KH}(\vmax)\equiv
{2\pttwo\zeta_{\rm max}\over\gamma(\gamma-1) }{\rs\over r_\pi}
\int_{\rs(\vmax)}^{r_\star(\vmax)}
{\delta\chi\over \M^2} {\d l\over r}\;. \label{AKH}
\end{equation}
The width $\grwdth$ of the entropy gradient decreases
geometrically with $r$, and the optimal wavelength of the \KHi
decreases according to Eq.~(\ref{EwaveKH}). By contrast,
the wavelength $\lambda$, parallel to the flow line, of a Lagrangian
perturbation must increase when advected, since the flow is accelerated.
It is therefore not possible to maintain the \KHi with its local maximum 
growth rate during the advection of the perturbation. 

If we denote by $\lambda_{\rm i}\equiv \seven h_{\rm i}\ll \rs$ the initial 
wavelength of the perturbation, Fig.~\ref{chi} indicates that it becomes 
unstable only when advected towards a region where the gradient width is 
$\grwdth(r)<2\lambda/\seven$. We rewrite Eq.~(\ref{AKH}) using 
Eq.~(\ref{delth}):
\begin{equation}
{\rm Min}\;{\cal A}_{\rm KH}=
{2\pttwo\zeta_{\rm max}\over\gamma(\gamma-1)}{\rs\over r_\pi}
\int_{\rs}^{r_\star} \left\lbrack
{\chi(\grwdth/\grwdth_{\rm i})\over\grwdth/\grwdth_{\rm i}}\right\rbrack
{1\over\M^2}{\grwdth_{\rm sh}\over\grwdth_{\rm i}}{\d l\over\rs}\;. 
\label{AKH2}
\end{equation}
Neglecting the increase of $\lambda_i$ due to the acceleration compared to 
the linear decrease of $h(r)$, the ratio 
$\lambda_{\rm max}(r)/\lambda_i\equiv\grwdth(r)/\grwdth_i$ decreases 
linearly to zero when $r\to0$. The integral in Eq.~(\ref{AKH2})
is approximated for $r_\star\to0$ by extracting some average values
$\hat\delta$ and $\hat\M$ from it, and integrating the function $\chi$ 
described in Fig.~\ref{chi}:
\begin{eqnarray}
\int_{0}^{\rs} \left\lbrack
{\chi(\grwdth/\grwdth_i)\over\grwdth/\grwdth_i}\right\rbrack
{1\over\M^2}{\grwdth_{\rm sh}\over\grwdth_i}{\d l\over\rs}
&\sim &{\hat\delta\over\hat\M^2}\int_0^2 {\chi(x)\over x}\d x,
\label{inte56}\\
&\sim &2.1\;{\hat\delta\over\hat\M^2} ,\label{estichi}
\end{eqnarray}
\begin{equation}
{\rm Min}\;{\cal A}_{\rm KH}\sim 
{0.4\over(\gamma-1)^2}\;{\pttwo\over 0.2}\;
{\zeta_{\rm max}\over0.15}\;
\hat\delta\;{\rs\over 2^{1\over2}r_\pi}
\left({\M_{\rm sh}\over\hat\M}\right)^2.
\label{AKH22}
\end{equation}
Introducing $\gamma=5/3$ into Eq.~(\ref{AKH22}), and approximating 
$\hat\M \sim (\M_{\rm sh}+1)/2$ we obtain:
\begin{equation}
{\rm Min}\;{\cal A}_{\rm KH}\sim 0.5 
\;{\pttwo\over 0.2}\;
{\zeta_{\rm max}\over0.15}\;{\hat\delta}\;
{\rs\over 2^{1\over2}r_\pi}.\label{effKH}
\end{equation}
We deduce from Eqs.~(\ref{entropapprox2}) and (\ref{effKH}) an estimate of 
the \KHi efficiency for $\gamma=5/3$:
\begin{equation}
0.5\la {\hat{\cal A}}_{\rm KH}\la 2.0\;.\label{AKHfinal}
\end{equation}
In particular, ${\hat{\cal A}}_{\rm KH}$ does not diverge when $r_\star\to0$.
According to Eq.~(\ref{inte56}), the integral in Eq.~(\ref{AKH2}) is regular 
at $r=0$, thus its value should not depend on the accretor size if 
$r_\star\ll r_\pi$. Consequently, the \KHi efficiency should be little 
influenced by the size of the accretor if $r_\star\ll r_\pi$.
The above calculation indicates that a whole range of wavelengths leads to
comparable efficiencies if $r_\star\ll r_\pi$. Although the integrated 
efficiency remains limited, the variability timescale is directly related
to the wavelength of the perturbation.
 \begin{figure}
%\picture 86.7mm by 72.3mm (max1)
\psfig{file=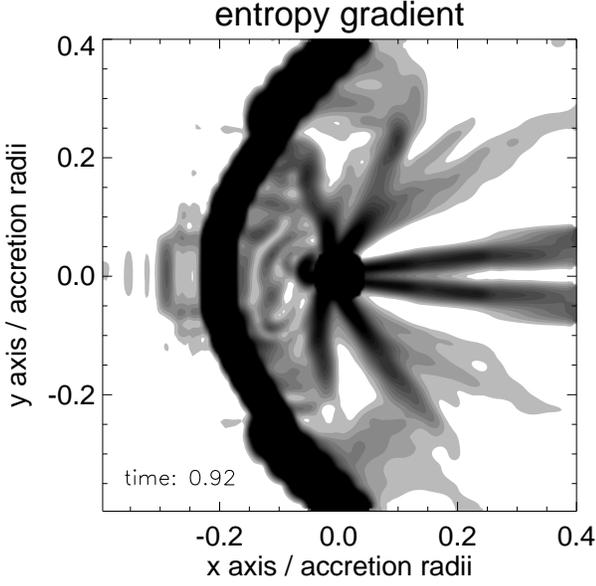,width=\columnwidth}
\caption[]{Splitting of the entropy gradient peak generated by the shock
into several narrower peaks.}   \label{maxvortsplit}
\end{figure}
Numerical simulations show that the gradient width, a priori comparable to
the radius of curvature of the shock, is split into several maxima of
much smaller width, typically of 5--10 degrees  
in a snapshot of the flow before instability (Fig.~\ref{maxvortsplit}). 
One may imagine that the gradient width decreases with
time, as more and more gas is accumulated from the downstream side of
the accretor, until the gradient width is short enough to start the
\KHi. We must also note that the generation of vorticity at the
interface between the grids in the multi--grid PPM numerical
technique might influence the distribution of vorticity in the flow.

\section{Discussion of the efficiencies of the two instabilities
\label{secdisc}}

The local time scale of the instability in the subsonic region scales like the 
advection time onto the accretor, so that the efficiency integrated 
along a flow line is always of order unity (Eqs.~\ref{ARTfinal} and \ref{AKHfinal}). 
Assuming that the threshold of the WKB approximation is also of the order unity
($\MA^*\sim 1$) and following the statements of Sect.~\ref{slocal}, 
we stand in the uncomfortable position between case (ii) and case (iii).
Despite this uncertainty, it is interesting to compare the analyze the results of 
numerical simulations in the light of these physical mechanisms.

The efficiency of the instabilities naturally increases with the strength of 
the shock, as can be seen from the function $\eta(\M_1)$ (Eq.~\ref{defeta} 
and Fig.~\ref{dS}). Numerical simulations with $\gamma=4/3$
(Ruffert 1995) and $\gamma=5/3$ (Ruffert \& Arnett 1994, Ruffert 1994b)
show the same trend: the flow is stable with $\M_\infty=1.4$ and unstable 
with $\M_\infty=3$ and $10$, with detached shocks in all these situations. 
The highest entropy gradients are obtained for 
$\M_\infty^2\ge 2/(\gamma-1)$, thus requiring higher Mach numbers for nearly 
isothermal flows. Increasing the Mach number does not increase the 
efficiency of the instabilities indefinitely: our estimates show that this 
efficiency saturates when the increase of the entropy gradient is compensated 
by the decrease of the advection timescale. We see no obvious justification 
for a possible divergence of any of the numerous dimensionless parameters 
introduced ($\zeta,\delta,g^*,\pttwo$) if the Mach number $\M_\infty$ is 
increased, or if the accretor size is decreased, and therefore expect
the integrated efficiencies to be smaller than $2$.

Our calculations seem to indicate that the efficiency should increase 
when $\gamma$ approaches unity (Eqs.~\ref{ART}, \ref{range} and \ref{AKH22}), 
whereas the most unstable flows observed in simulations correspond to 
$\gamma=4/3$ and $\gamma=5/3$. Nevertheless, we showed in 
Sect.~\ref{secattached} that the efficiency of the \RTi is finite
for $\gamma=1$ when the shock is attached to the accretor. Thus 
this paradox would disappear if a critical adiabatic index 
exists below which the shock is attached to the accretor, as suggested by 
Wolfson (1977). 

The axisymmetric \KHi must be considered together with the effect 
of stratification, which can be stabilizing or destabilizing depending on 
the sign of $\beta$. The influence of stratification on the \KHi can 
be estimated quantitatively by comparing the time scales associated to 
each physical process, in the same spirit as was done with the Richardson 
number. This ratio informs us about a possible stabilization of the \KHi by 
buoyancy forces if $\beta<0$, or which of the two instabilities is the 
fastest if $\beta>0$. The ratio of the timescales can be deduced from 
Eqs.~(\ref{sRT}), (\ref{ErateKH}) and (\ref{vormax}) as follows:
\begin{equation}
{\growth_{\rm RT}\over \growth_{\rm KH}}=
{1\over\pttwo}\left\lbrack{\gamma(\gamma-1)\over2}g^*\right\rbrack^{1\over2}
\M^2{v_{\rm ff}\over v} {\sin^{1\over2}\beta\over\chi}
\left( r\nabla S_{\rm max}\right)^{-{1\over2}}.\label{ratio}
\end{equation}
We estimate this ratio both immediately after the shock and at the sonic
point. At the shock, the effective gravity is strong ($g^*\sim1$), and 
Eqs~(\ref{entropapprox2}), (\ref{range}) and (\ref{mach2}) provide us with a 
lower bound for Eq.~(\ref{ratio}):
\begin{eqnarray}
{\growth_{\rm RT}\over \growth_{\rm KH}}(\rs)&\ge &
3.8 (\gamma-1)^{3\over2}\left({\gamma+1\over\gamma}\right)^{1\over2}
{\sin^{1\over2}\beta\over\chi}
\nonumber\\
&\times&\left({0.15\over\delta\zeta_{\rm max}}\right)^{1\over2}
{0.2\over\pttwo}\left({2^{1\over2}r_\pi\over \rs}\right)^{1\over2}.
\label{ratiosg}
\end{eqnarray}
Applying this equation to $\gamma=5/3$, we obtain:
\begin{equation}
{\growth_{\rm RT}\over \growth_{\rm KH}}(\rs)\ge 2.6
\;{\sin^{1\over2}\beta\over\chi}
\left({0.15\over\delta\zeta_{\rm max}}\right)^{1\over2}
{0.2\over\pttwo}\left({2^{1\over2}r_\pi\over \rs}\right)^{1\over2}.
\label{ratiosg}
\end{equation}
Since $\beta$ is large immediately after the shock, 
Eq.~(\ref{ratiosg}) suggests that the \RTi is more unstable than the \KHi 
there. At the sonic point ($\M=1$), using Eqs.~(\ref{entropapprox2}) 
and (\ref{range}), Eq.~(\ref{ratio}) becomes:
\begin{eqnarray}
{\growth_{\rm RT}\over \growth_{\rm KH}}(r_{\rm sonic})&\ge &
6.6 \gamma^{1\over2}(\gamma-1)
{\sin^{1\over2}\beta\over\chi}
\nonumber\\
&\times&\left({g^*\over 0.75}\right)^{1\over2}
\left({0.15\over\delta\zeta_{\rm max}}\right)^{1\over2}
{0.2\over\pttwo}\left({2^{1\over2}r_\pi\over \rs}\right)^{1\over2}.
\label{ratiosonic}
\end{eqnarray}
Applying this equation to $\gamma=5/3$, we obtain:
\begin{eqnarray}
{\growth_{\rm RT}\over \growth_{\rm KH}}(r_{\rm sonic})&\ge&
5.7 \;{\sin^{1\over2}\beta\over\chi}
\nonumber\\
&\times&\left({g^*\over 0.75}\right)^{1\over2}
\left({0.15\over\delta\zeta_{\rm max}}\right)^{1\over2}
{0.2\over\pttwo}\left({2^{1\over2}r_\pi\over \rs}\right)^{1\over2}.
\label{ratiosonicg}
\end{eqnarray}
Let us first remark from Eq.~(71) of Paper~I that if $\gamma=5/3$, 
$0\le\lim_{r\to 0}(\p\log\beta/\p\log r)\le1$, indicating 
that the convergence of $\sin^{1\over2}\beta$ towards zero is slower than 
that of $\chi(r)\sim r/2$ close to a point like accretor. Thus the \KHi
ultimately becomes stabilized by buoyancy forces near a point like accretor,
when the wavelength of the perturbation is much longer than the vorticity
gradient width.  

According to Eq.~(\ref{ratiosonicg}), a perturbation with a wavelength such 
that $\chi=1$ at the sonic point would be locally unstable to the \KHi with 
hardly any stabilization by the buoyancy forces if $|\beta |\la 1$ degree. 
The angle $\beta$ at the sonic point is negative for $\gamma=5/3$ 
(see Paper~I) and much smaller than at the shock, especially if the sonic 
surface is close to the accretor. We know from Paper~I that the sonic surface 
is attached to the accretor if $\gamma=5/3$. Thus the flow line reaching a 
point like accretor along the sonic surface is unstable to the \KHi without 
being stabilized by the buoyancy forces. The shortest timescale associated 
to the \KHi therefore depends on the size of the accretor, while the longest 
is directly related to the shock distance.

Although the \RTi is dominant near the shock, and the \KHi might prevail 
closer to the sonic surface, it is hard to disentangle the two mechanisms, 
which act in a combined way when the shock is detached. The situation is 
different when the shock is attached to the accretor, since there is no
vorticity maximum in this case. It is interesting to note 
that the linear efficiency estimated in Eq.~(\ref{attached}) for an 
attached shock is not much smaller than the efficiencies derived for 
a detached shock. Numerical simulations indicate that 
accretion flows with an attached shock are generally more stable than those 
with a detached shock (Ruffert 1995, Zarinelli \etal 1995). 
In the simulations by Ruffert (1996) with $\gamma=1.01$, the instability is 
nevertheless clearly observed, producing fluctuations of the mass accretion 
rate of about $7\%$ (simulations with $\gamma=4/3$ and $5/3$ produce 
typical fluctuations of $14\%$ and $23\%$ respectively). 
This suggests that the \RTi alone may destabilize 
linearly the flow when the shock is attached, while the combined action 
of the two mechanisms when the shock is detached is more efficient and 
leads to higher non linear amplitudes.

\section{Instabilities in simulations with subsonic flows\label{snew}}
\begin{figure*}
  \begin{tabular}{cc}
    \psfig{file=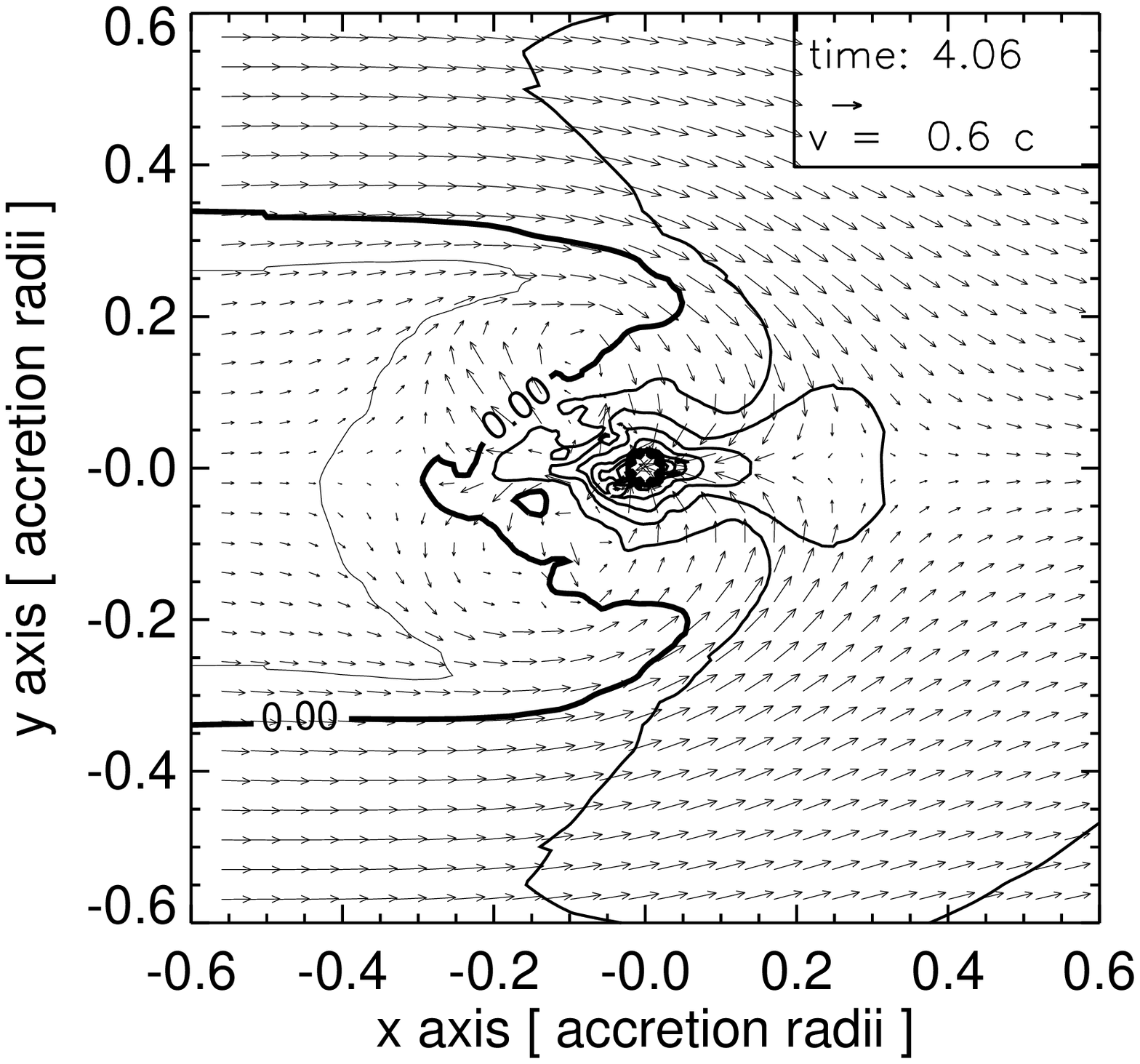,width=\columnwidth} &
    \psfig{file=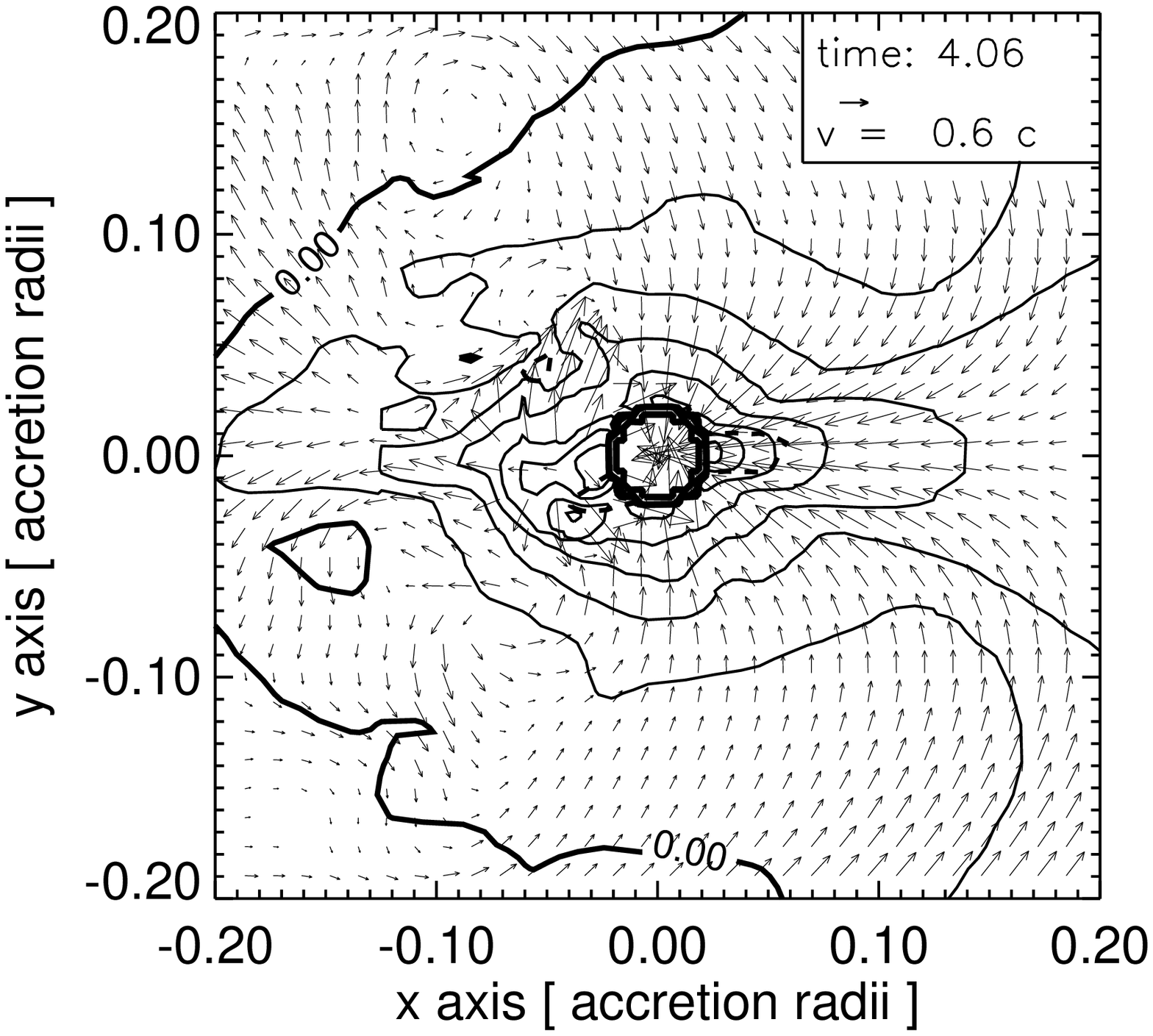,width=\columnwidth} \\
  \end{tabular}
\caption[]{Contour plot of the density distribution together with the
instantaneous flow velocities.
The contours are spaced logarithmically with intervals of 0.1 dex,
the contour corresponding to $10^0$ is annotated. The time elapsed
since the beginning of the simulation together with the velocity
arrow unit is shown at the top right. The accretor is at the center
with coordinates $x=0, y=0$. The right figure is a zoom of the left
figure enlarged around the accretor.
\label{fig:contour}
}
\end{figure*}
In order to separate more clearly the different effects that might
contribute to the formation of the instabilities, we performed
numerical simulations of accretion from a flow with subsonic bulk
velocities at infinity, but with a gradient in entropy. The rationale
is the following.
When the accretor moves supersonically relative to a surrounding
medium a bow shock will form.
The shock transforms the initially (at infinity) homogeneous medium
moving at supersonic velocities into a subsonic flow with an entropy
decreasing away from the axis of symmetry.

We did a set of simulations to mimic these conditions, of which we
will present here the most
important results relevant to the topic of this paper. The accretor
radius is chosen to be $r_\star=0.02 \rA$ and
the same absorbing boundary conditions are used as in Ruffert (1994a).
The simulations are done on 7~nested cartesian grids, the zone size
on each finer grid being a factor of two smaller than of the next
coarser grid.
The largest grid spans 8 accretion radii. Matter is evolved
hydrodynamically using the ``Piecewise Parabolic Method'', assuming it
is a polytropic gas with an adiabatic index of~5/3.

The flow at infinity is in direction of the positive $x$-axis, with a
constant velocity along $x$. The pressure is uniform at infinity. In
order to maintain pressure equilibrium, the maximum of entropy 
$\Delta S$ at $y=0$ at infinity along the $x$-axis is offset by a 
minimum in density with the following shape:
\begin{eqnarray}
\frac{\rho_\infty(x,y,z)}{\rho_{\rm 0}} &\equiv &
1- \varepsilon \exp\left(-\frac{y^2}{2}\right)\;, \label{eq:dens}\\
{\rm with} \;\varepsilon &\equiv&
1-\exp\left\lbrack-{\gamma-1\over\gamma} \Delta
S\right\rbrack \;. 
\end{eqnarray}
The entropy difference was chosen to be $\Delta S = 2$ 
($\varepsilon=0.55$, $\rA\nabla S_{\rm max}\sim 1.33$) which 
is comparable to the value observed in the numerical simulations, 
e.g.~in Fig.~17d of Ruffert (1994b) and Fig.~\ref{vortflow} of 
this paper. The velocity at infinity was chosen constant, such 
that the Mach number at infinity is 0.6 far from the axis. The 
density profile (Eq.~\ref{eq:dens}) and the uniform pressure
imply that the Mach number decreases by a factor
$(1-\varepsilon)^{1/2}$ along the axis.

Fig.~\ref{fig:contour} shows a snapshot of the density and velocity
distribution of such a model. At the point in time at which the
snapshot was taken two transient vortices are prominent at
($x\approx-0.1,y\approx\pm0.15$), corresponding to a buoyant
vortex ring rising against the flow. Such rings were already observed in
numerical simulations by Koide, Matsuda \& Shima (1991, Fig.~12).
When boundary conditions were chosen for which there is little or no 
accretion at the surface of the star, \eg by  Shima \etal
(1985), Fryxell, Taam \& McMillan (1987), Matsuda \etal (1991),
such vortex structures appeared too.

A movie showing the temporal
evolution of the density distribution shows that the flow is not
stationary. Just from an inspection of the contour plots and movies
no obvious difference is apparent between the instabilities generated
in these simulations and the instabilities present in the simulations
of three-dimensional BHL--accretion (e.g. Ruffert~1996, and references
therein).

Since the entropy gradients are present over the whole domain of the 
simulation, it is important to check the contribution of the region
far from the accretor (\eg $r>\rA$) to the efficiency of the
instabilities. Thus we decompose the global efficiency of each
instability into the contributions ${\cal A}_i$ and ${\cal A}_i^{\infty}$ 
defined as follows:
\begin{eqnarray}
\int_{r_\infty}^{r_{*}}{\growth_{i}\over v}\d l
 &=& \int_{r_\infty}^{\rA}{\growth_{i}\over v}\d l
+\int_{\rA}^{r_{*}}{\growth_{i}\over v}\d l,\\
&\equiv & {\cal A}_i^\infty + {\cal A}_i
\end{eqnarray}
Far ahead from the accretor, the trajectories are nearly parallel to the 
symmetry axis, and the growth rate of the \RTi along a flow line
indexed by $\varpi\sim \rA$ can be estimated from 
Eq.~(\ref{sRT}) with $\sin\beta\sim \varpi/r$:
\begin{eqnarray}
\growth_{\rm RT}&=&{\gamma-1\over\gamma} \varpi\nabla S
{GM\over r^{3\over2}},\\
{\cal A}_{\rm RT}^{\infty}&<& 
\left({\gamma-1\over2\gamma} \rA\nabla S\right)^{1\over2}
\int_{r_\infty}^{\rA\over\varpi} {\d u\over (1+u^2)^{3\over4}},\\
&<&0.6,
\end{eqnarray}
where  we have used the parameters of the simulation, namely 
$r_\infty/\rA=8$, and $\gamma=5/3$.
The convergence of the integral when $r_\infty\to\infty$ ensures that 
the main contribution of the \RTi comes from the region close to the 
accretor, within a few accretion radii. Nevertheless, the value of 
${\cal A}_{\rm RT}^{\infty}$ might not be fully negligible compared to
${\cal A}_{\rm RT}$ as estimated in Eq.~(\ref{ARTfinal}).

An important difference between this model and the supersonic models
with a bow shock is that here the Ber\-noul\-li constant
$B(r,\theta)$, defined by $B\equiv v_\infty^2/2+c_\infty^2/(\gamma-1)$
is not uniform far away from the accretor, since the sound speed is not
uniform at infinity. The non--uniformity of $B$ adds an additionnal term
in Eq.~(\ref{vortentro}) (Eq.~9 of Paper~I):
\begin{equation}
{\vec w}\times {\vec v}=T{\vec \nabla}
S-{\vec\nabla}B.\label{vortbern} 
\end{equation}
Using the property that the entropy and the Bernoulli constant are conserved
along the stationary flow lines, and that the vorticity is zero at infinity,
we deduce from Eq.~(\ref{vortbern}) that the gradient of $B$ is 
proportionnal to the entropy gradient and rewrite the vorticity as follows:
\begin{eqnarray}
\nabla B&=&{c_\infty^2\over\gamma}\nabla S,\\
w&=&{\nabla S\over \gamma v}(c^2-c_\infty^2).
\label{vortib}
\end{eqnarray}
The increase of the temperature close to 
the accretor therefore implies that the contribution of the term 
${\vec\nabla} B$ in Eq. (\ref{vortbern}) becomes negligible close to the 
accretor, like in BHL flows.

The Bernoulli equation (\ref{Bernoulli}) is used to obtain an upper bound 
for the variation of the sound speed far from the accretor:
\begin{eqnarray}
c^2-c_\infty^2&=&{(\gamma-1)c_\infty^2\over 2 + (\gamma-1)\M^2}
\left\lbrack{2GM\over rc_\infty^2} - (\M^2 - \M_\infty^2)\right\rbrack,\\
&<&{\gamma-1\over 2 + (\gamma-1)\M^2}
\left({2GM\over r}\right).
\end{eqnarray}
Using Eqs.~(\ref{ErateKH}) and (\ref{vortib}) with optimal wavelength and 
neglecting compressibility effect, we obtain the following upper bound for 
the contribution of the region ahead of the accretor to the efficiency of 
the \KHi:
\begin{eqnarray}
{\cal A}_{\rm KH}^\infty
&<& \alpha {\gamma-1\over 2 + (\gamma-1)\M_\infty^2}
{\rA\nabla S\over \gamma}
\log\left({r_\infty\over \rA}\right),\\
&<&0.07,
\end{eqnarray}
where $r_\infty/\rA=8$, $\M_\infty=0.6$, and $\gamma=5/3$.

Thus both ${\cal A}_{\rm RT}^\infty$ and ${\cal A}_{\rm KH}^\infty$
are smaller than the estimated values of ${\cal A}_{\rm RT}$  
(Eq.~\ref{ARTfinal}) and ${\cal A}_{\rm KH}$ (Eq.~\ref{AKHfinal}) in the 
vicinity of the accretor, although they might not be fully negligible.

We conclude that these simulations , which
might possibly at first sight seem artificial, are an encouraging
indication of the instabilities being generated by entropy gradients
in the flow. They show that an instability exists which does not rely on 
the deformations of the shape of the shock surface, nor on the reflection 
of waves against the shock surface. 
Moreover, they suggest that the instability of the BHL flow is
not an artifact due to the numerical treatment of the shock.

These simulations should be followed by other
simulations in order to understand better this instability. A first step 
would be to explore numerically the effect of the amplitude of the 
entropy gradient. The absence of shocks makes this flow easier to
study by a global perturbation analysis in order to obtain conclusive 
statements about the precise onset of linear stability of the flow. 
Particular analytical solutions of such flows would be very useful in this 
respect.

\section{Conclusions \label{Sconclusion}}

Despite our lack of knowledge concerning both the shape of the shock 
surface, and the dependence of the shock distance on the adiabatic index 
and Mach number, we are able to make a quantitative estimate of the entropy
gradients produced by the shock. A priori, entropy gradients and vorticity are
sources of linear instability through the Rayleigh--Taylor and the 
Kelvin--Helmholtz mechanisms. 
Their efficiency is estimated using a WKB like analysis, by integrating their
local linear growth rate along a flow line between the shock and the accretor.
The WKB criterion is only marginally satisfied, since the growth time is at 
best comparable to the advection time onto the accretor. This may cast doubts 
on the significance of our quantitative estimates, although the physical 
picture seems robust. If correct, this suggests that only large enough initial 
perturbations may reach non linear amplitudes when amplified by these 
mechanisms. 

It is striking that several features of the instability observed in 
numerical simulations (cf.~the animations currently available at 
{\tt http:\nix//www.mpa-garching.mpg.de\nix/
\lower0.7ex\hbox{$\!$\~{}}mor\nix/bhla.html})
would be explained naturally by these mechanisms:
\par (i) {\it the instability requires a shock}: both the \RTi and 
the \KHi require the presence of a shock, which produces entropy 
gradients and vorticity.
\par (ii) {\it the instability is stronger when the shock is detached}:
the \KHi adds to the destabilization of the flow if the shock is detached.
\par(iii) {\it the instability is stronger for high Mach numbers}: the
entropy gradient and vorticity produced by the shock increase with the Mach 
number.
\par(iv) {\it the instability is stronger for small accretors}: 
the smaller the accretor, the longer the advection time, and the
stronger the entropy gradients.
\par(v) {\it the instability is nonaxisymmetric}: as is the \RTi.
\par(vi) {\it the instability starts in the region of intermediate azimuthal
angle ($\theta\sim \pi/2$), close to the accretor}: this region coincides
with the region of maximum entropy gradient and vorticity, if the shock is
detached.\\

If these mechanisms were responsible for the instability, our
calculations further indicate that the efficiency of the linear instability 
becomes 
\par (i) independent of the Mach number if the kinetic energy dominates the
internal energy: $\M_\infty^2\gg 2/(\gamma-1)$,
\par(ii) independent of the accretor size if it is much smaller than the 
shock distance: $r_\star\ll r_\pi$. Nevertheless, the range of timescales
associated to the instability depends naturally on the size of the 
accretor.\\

A higher efficiency of the linear instabilities would be reached if a 
feedback loop could be obtained: this is the subject of ongoing research. 

Our study should invite the groups performing numerical simulations to
follow carefully the evolution of entropy gradients and vorticity, since 
an overestimate of these quantities could lead to an artificially 
strong instability.

The simulations presented in Sect.~\ref{snew} suggest that an absorbing
accretor moving with a {\it subsonic} velocity in a gas with a non uniform 
entropy gives rise to an unstable accretion flow, resembling the BHL 
instability. This configuration might be a fruitful approach to 
understand the BHL instability better.

\acknowledgements
Numerical simulations and investigations were accomplished in the
productive environment of the Max Planck Institut f\"ur Astrophysik,
Garching, Germany. TF was supported by the EC grant
ERB-CHRX-CT93-0329, as part of the research network 'accretion onto
compact objects and protostars'. TF thanks the IoA for kind hospitality.
MR thanks PPARC for support and the
SAp for kind hospitality. The authors are grateful to M. Tagger for 
insightful discussions. We would like to thank the referee Dr. U. Anzer 
for a prompt and thorough review of this paper.

\appendix

\section{The Kelvin--Helmholtz and Rayleigh--Taylor instabilities in
a compressible fluid \label{AKHRT}}

\subsection{Stratified sheared atmosphere\label{aprtsh}}

We consider a stratified atmosphere with a density profile $\rho_0(z)$,
entropy $S_0(z)$, sound speed $c_0(z)$ in a vertical constant gravity $g_0$ 
satisfying Eq.~(\ref{equili}). The atmosphere is sheared in the 
$y$--direction with the velocity profile $V_0(z)$.
Perturbations are decomposed on a Fourier basis in the $x,y$ directions,
with their associated wavenumbers $k_x,k_y$. Their growth rate is denoted by
$\growth$.
The linearized equations for perturbations lead to the following 
differential system on the perturbed vertical velocity $v_z$ and pressure 
$p$:
\begin{eqnarray}
{\p(i\rho_0v_z)\over\p z}&=&
\left\lbrack {k_yV_0'\over k_yV_0-i\growth}-{\gamma-1\over\gamma}
{\p S_0\over \p z} \right\rbrack
(i\rho_0v_z)\nonumber\\
& &+\left\lbrack \left({k_yV_0-i\growth\over c_0}\right)^2-k_x^2-k_y^2
\right\rbrack {p\over k_yV_0-i\growth}\label{aprt1}\\ 
{\p p\over \p z}&=&-\left\lbrack
(k_yV_0-i\growth)^2-{\gamma-1\over\gamma}g_0{\p S_0\over \p z}
\right\rbrack {(i\rho_0v_z)\over k_yV_0-i\growth}
\nonumber\\
& &- {g_0\over c_0^2}p\;.\label{aprt2}
\end{eqnarray}
Taking the limit $k_x\to\infty$, $k_y=0$ in
Eqs.~(\ref{aprt1})-(\ref{aprt2}), we
recover the Brunt--V\"ais\"al\"a frequency defined in Eq.~(\ref{defsrt}).

\subsection{Compressible \KHi without stratification\label{akhc}}

The effect of compressibility on the linear \KHi can be
studied by solving Eqs.~(\ref{aprt1})--(\ref{aprt2}) for $g_0=0$.
The pressure $P_0(z)$ across the flow is taken constant for equilibrium,
but the
entropy and the temperature may vary.\\
$V_0(z)$ being the unperturbed flow speed in the $y$--direction, we
obtain the following differential equation for the pressure
perturbation:
\begin{eqnarray}
{\p^2 p\over\p z^2}&-&2{\p\log\over\p z}\left({k_yV_0-i\growth\over
c_0}\right) {\p p\over\p z}\nonumber\\
&+&\left\lbrack\left({k_yV_0-i\growth\over c_0}\right)^2
-k_x^2-k_y^2\right\rbrack p =0\;.
\label{sommerfeld}
\end{eqnarray}
The Orr--Sommerfeld equation governing the incompressible case 
(\eg Drazin \& Reid 1981) is
recovered by taking the limit $c_0\to\infty$. We solved this equation
by a relaxation method between two rigid walls, and obtained 
Figs.~\ref{chi} and \ref{figmach}. Different entropy profiles were tested,
particularly the ones keeping a uniform Bernoulli constant with a
length scale comparable to the vorticity length scale (as in an
entropy--induced vorticity), with no significant departure from the
case of uniform entropy. Indeed, we see from equation
(\ref{sommerfeld}) that a non uniform entropy does not introduce
significant changes compared to the case of uniform entropy, apart
from the necessary spatial dependence of the sound speed $c_0(z)$.

\subsection{Combined actions of shear and buoyancy}

The Richardson number Ri measures the ratio of buoyancy force to
inertia for an incompressible fluid (see for example
Chandrasekhar, 1961, \S103):
\begin{equation}
{\rm Ri}\equiv-g_0\left({\p\log\rho\over\p z}\right) \left({\p
V_0\over\p z}\right)^{-2} \;.
\end{equation}
It is also the squared ratio of the time scales associated to the 
\RTi and \KHi.
In the classical case of an exponential density and a hyperbolic
tan-velocity profile, the Kelvin--Helmholtz instability is
stabilized for ${\rm Ri}>1/4$.

\section{Table of symbols used\label{Ap2}}
In Table \ref{table} we present a list of the most important symbols used 
in this paper together with the equation number of their first occurence.

\begin{table}[h]
\caption[]{Equation number corresponding to the first occurence of the 
symbols used}
\label{table}
\begin{tabular}{ccl}
\hline\\[-3mm]
symb.  & Eq. & comment \\
\hline\\[-3mm]
$r$, $\theta$   &       
& spherical coordinates \\
$\rA$     &  
& accretion radius      \\
$\rs$		& \ref{defdelta}
& shock shape \\
$r_\pi$		& \ref{deltamax}
& shock distance along the sym. axis\\
$r_\beta(\theta)$       & \ref{ERTgeneral}        
& surface where the flow is radial\\
$c$     &       
& sound speed   \\
$c_0$   & \ref{critcomp} 
& sound speed at the vorticity maximum \\
$w$     & \ref{vortentro}
& vorticity    \\
$v$     & \ref{vortentro}  
& gas velocity  \\
$v_{2\perp}$	& \ref{ds2}
& velocity after and perp. to the shock\\
$\vff$		&\ref{vff}
& free fall velocity\\
$g_{\rm eff}$   &    \ref{defgeff}   
& effective gravity     \\
$g^*$   &    \ref{defg*}   
& dimensionless gravity     \\
$h$     & \ref{EwaveKH} 
& vorticity gradient width        \\
\hline
$G$     &       
& gravitational constant \\
$M$     &       
& accretor mass \\
$S$     & \ref{vortentro}
& entropy      \\
$T$     & \ref{vortentro}
& temperature  \\
$L$		& \ref{ds2}
& curvilinear abscisse along the shock\\
\hline
$\M_\infty $      &       
& mach number at infinity   \\
$\M_1$    &\ref{entroRH} 
& mach number perp. to and before the shock        \\
$\M_{\rm sh}$ & \ref{mach2}
& mach number immediately after the shock \\
${\cal M}_{\rm KH}$ & \ref{critcomp} 
& relative mach number \\
${\cal A}(\varpi)$      & \ref{EcalA} 
& strength of instability along a flow line\\
${\hat{\cal A}}$      & \ref{defma} 
& maximum strength of instability\\
\hline
$\gamma$        & \ref{entroRH} 
& adiabatic index        \\
$\growth $      & \ref{WKB} 
& linear growth rate      \\
$\varpi$	& \ref{EcalA}
& flow line indexed by its impact parameter\\
$\eta$  & \ref{defeta} 
& dimensionless entropy jump strength \\
$\theta_{\rm sh}$	& \ref{v1pe}
& opening angle of the attached shock\\
$\theta_{\rm so}$	& \ref{attached}
& opening angle of the attached sonic surface\\
$\zeta$         & \ref{defzeta}      
& dimensionless entropy gradient\\
$\zeta^*$	& \ref{defzetastar}
& min. value of the max. entropy gradient\\
$\delta$        & \ref{defdelta} 
& dimensionless geometrical factor\\
$\beta$ & \ref{defbeta} 
& flow line angle       \\
$\pttwo$        & \ref{ErateKH} 
& \KHi geometrical factor      \\
$\seven$        & \ref{EwaveKH} 
& \KHi optimal wavelength       \\
$\chi$  & \ref{ErateKH} 
& \KHi wavelength dependency \\
$\lambda$       & \ref{EwaveKH} 
& wavelength of the perturbation\\
\hline
\end{tabular}
\end{table}

\end{document}